\documentclass[11pt]{article} % use larger type; default would be 10pt

\usepackage[utf8]{inputenc} % set input encoding (not needed with XeLaTeX)

\usepackage{geometry} % to change the page dimensions
\usepackage{graphicx} % support the \includegraphics command and options

\usepackage{booktabs} % for much better looking tables
\usepackage{array} % for better arrays (eg matrices) in maths
\usepackage{paralist} % very flexible & customisable lists (eg. enumerate/itemize, etc.)
\usepackage{verbatim} % adds environment for commenting out blocks of text & for better verbatim
\usepackage{subfig} % make it possible to include more than one captioned figure/table in a single float
\usepackage{amsmath}
\usepackage{amssymb}
\usepackage{enumerate}
\usepackage{natbib}
\usepackage{url} % not crucial - just used below for the URL 
\usepackage{bm}
\usepackage{multirow}
\usepackage{booktabs}
\usepackage{bbm}
\usepackage{tikz}
\usetikzlibrary{shapes.arrows,decorations.pathreplacing}
\usepackage[english]{babel}

\usepackage{fancyhdr} % This should be set AFTER setting up the page geometry
\usepackage{sectsty}
\usepackage{setspace}
\usepackage[nottoc,notlof,notlot]{tocbibind} % Put the bibliography in the ToC
\usepackage[titles,subfigure]{tocloft} % Alter the style of the Table of Contents

\newcommand{\by}{\mathbf{y}}

\newcommand{\bX}{\mathbf{X}}
\newcommand{\bx}{\mathbf{x}}
\newcommand{\bbeta}{\boldsymbol\beta}

\newcommand{\bD}{\mathbf{D}}

\newcommand{\bZ}{\mathbf{Z}}
\newcommand{\bd}{\mathbf{d}}
\newcommand{\bdelta}{\boldsymbol{\delta}}
\newcommand{\bepsilon}{\boldsymbol{\epsilon}}
\newcommand{\blambda}{\boldsymbol{\lambda}}
\newcommand{\bgamma}{\boldsymbol\gamma}

\newcommand{\bSigma}{\boldsymbol\Sigma}

\newcommand{\bA}{\mathbf{A}}
\newcommand{\bK}{\mathbf{K}}

\newcommand{\vbar}{\,|\,}

\newcommand{\E}[2]{ {\mathbb E}_{#1} \left[ #2 \right] }

\newcommand{\Exx}{\mathbb{E}}

\DeclareMathOperator*{\argmin}{arg\,min}

\addto{\captionsenglish}{%
  
}

\title{\bf Bayesian Sparse Global-Local Shrinkage Regression for Selection of Grouped Variables}
\author{Zemei Xu\thanks{Melbourne School of Population and Global Health, School of Mathematics and Statistics, University of Melbourne (Email: zemeix@student.unimelb.edu.au)} \and 
Daniel F. Schmidt \thanks{Melbourne School of Population and Global Health, University of Melbourne} \and 
Enes Makalic \footnotemark[2] \and
Guoqi Qian \thanks{School of Mathematics and Statistics, University of Melbourne} \and 
John L. Hopper \footnotemark[2]}

\begin{document}
\maketitle

\begin{abstract}
Most estimates for penalised linear regression can be viewed as posterior modes for an appropriate choice of prior distribution. Bayesian shrinkage methods, particularly the horseshoe estimator, have recently attracted a great deal of attention in the problem of estimating sparse, high-dimensional linear models. This paper extends these ideas, and presents a Bayesian grouped model with continuous global-local shrinkage priors to handle complex group hierarchies that include overlapping and multilevel group structures. As the posterior mean is never a sparse estimate of the linear model coefficients, we extend the recently proposed decoupled shrinkage and selection (DSS) technique to the problem of selecting groups of variables from posterior samples. To choose a final, sparse model, we also adapt generalised information criteria approaches to the DSS framework. To ensure that sparse groups, in which only a few predictors are active, can be effectively identified, we provide an alternative degrees of freedom estimator for sparse Bayesian linear models that takes into account the effects of shrinkage on the model coefficients. Simulations and real data analysis using our proposed method show promising performance in terms of correct identification of active and inactive groups, and prediction, in comparison with a Bayesian grouped slab-and-spike approach.
%200 or fewer words.
\end{abstract}

\noindent%
\vfill

\newpage

%===================== Section =====================%
\setstretch{1.3} % Line spacing of 1.3
\section{Introduction}
\label{sec:intro}

Consider the standard Gaussian linear regression model
\begin{align*}
	\by = \mu \mathbf{1}+ \bX\bbeta +\bepsilon,
\end{align*} 
where $\by = (y_1,\ldots,y_n)^T$ are observations of a response variable, $\bX = (\bx_1^T,\ldots,\bx_n^T)^T$ is an $n \times p$ observation matrix of the predictor variables, $\mu$ is the intercept, $\bbeta = (\beta_1,\ldots,\beta_p)^T$ is a vector of regression coefficients to be estimated from the data and $\bepsilon = (\epsilon_1,\ldots,\epsilon_n)^T$ are independent Gaussian random disturbances. 

In recent years, estimation of high-dimensional regression models for big data has become a prominent problem in many fields of research. %There is an increasing demand for advanced statistical methods to deal with the problems associated with big data. In this case, 
Here, the number of regression coefficient parameters to be estimated, $p$, is assumed to be of high dimension and often exceeds the number of observations, $n$. A further assumption when analysing high-dimensional data by a regression model is that most of the predictors are unassociated with the response variable. In other words, the vector $\bbeta$ is assumed to be sparse. 

Conventional approaches to estimating regression coefficients, such as least-squares and method of maximum likelihood, 
%which minimise a function that measures how well the estimated model fits the observed data. However, these two methods 
may break down and perform poorly when applied to high-dimensional data. Penalised regression methods, especially those based on $\ell_1$ regularisation, such as the Lasso \citep{1996Tibshiranip267288}, 
%are popular alternatives to maximum likelihood. Penalised regression methods work by shrinking regression coefficients to reduce overfitting and remove model indeterminacy. 
provide popular alternatives to tackle these difficulties. As an example, the Lasso is able to perform both shrinkage and variable selection by minimising a goodness of fit term plus a penalty proportional to the sum of the absolute values of the regression coefficients estimates. This may force some of the coefficients to be estimated as exactly zero; therefore the Lasso can produce sparse estimates.

Most penalised regression methods can be viewed as Bayesian procedures by interpreting the estimate of $\bbeta$ as the posterior mode under an appropriate prior distribution. For example, the Lasso procedure can be viewed as maximum a posteriori estimation under a Gaussian linear regression model when the coefficients $\bbeta$ follow independent and identical double exponential prior distributions; this is known as the Bayesian Lasso \citep{2008Parkp681686}. 

%Under the Bayesian framework, there are two broad approaches to sparse estimation and variable selection: 
Two broad classes of priors are commonly used in Bayesian sparse estimation and variable selection: (1)~two-component finite mixture priors, and (2)~continuous shrinkage priors. A two-component finite mixture prior, also known as a spike-and-slab prior \citep{1988Mitchellp10231032,1993Georgep881889}, consists of a point mass at $\beta_j=0$ and an absolute continuous alternative density for $\beta_j \neq 0$. Although the two-component finite mixture prior can produce estimates of regression coefficients exactly equal to zero, it entails an exploration of a model space of size $2^p$; computational difficulties therefore arise as the number of predictors $p$ grows. 

For the continuous shrinkage prior approach, each $\beta_j$ is assigned a continuous shrinkage prior centered at $\beta_j = 0$. One of the most important classes of continuous shrinkage priors are the global-local shrinkage priors \citep{2010Polsonp501538} whereby the prior distribution for $\bbeta$ can be expressed using the hierarchy
\begin{align*}
	\beta_j \,|\,\lambda_j^2,\tau^2 &\,\sim\, N(0,\tau^2\lambda_j^2),\\
	\lambda_j^2 &\,\sim\, \pi(\lambda_j^2), \\
	\tau^2 &\,\sim \,\pi(\tau^2).
\end{align*}
Here, $\lambda_j$ is the local shrinkage parameter associated with the $j$th predictor, $\tau$ is the global shrinkage parameter that controls the overall degree of shrinkage, and $\pi(\cdot)$ are generic prior distributions or density functions, and do not necessarily refer to the same function in different occasions. The particular choice of the prior distribution for the local shrinkage parameters $\bm{\lambda} = (\lambda_1,\ldots,\lambda_p)$ determines the behaviour of the resulting Bayesian estimator. This global-local hierarchy includes many important models such as the horseshoe~\citep{2010Carvalhop465480}, the horseshoe+~\citep{2016Bhadrap} and the Dirichlet--Laplace~\citep{2015Bhattacharyap14791490} etc. This approach offers great computational advantages in comparison with the two-component spike-and-slab mixture priors: there is no requirement to explore a discrete model space of size $2^p$. However, one limitation of the continuous shrinkage prior approach is that it fails to provide a sparse solution. To address the problem, specialised methods for producing sparse estimates from posterior samples, such as penalised variable selection based on posterior credible regions (\cite{2012Bondellp16101624}) or the decoupled shrinkage and selection method~(\cite{2015Hannp435448}), have been developed.

%%From a Bayesian perspective, the Bayesian model can be constructed. Without loss of generality, the response variable is centered and the covariates are standardised to have zero mean and unit length:
%
%\begin{align*}
%	\by = \by-\mu_{\by},\ \sum\limits_{i=1}^n x_{ij}/n=0, \ \sum\limits_{i=1}^n x_{ij}^2/n=1,\mbox{ for }j=1,\cdots,p,
%\end{align*} 
%
%where $\mu_{\by}$ is the mean of response variable. 

Without loss of generality, we assume that the response variable is centered and the covariates are standardised to have a zero mean and unit variance. 
%This convention is also adopted throughout this paper. 
A general hierarchical Bayesian global-local shrinkage regression model can be expressed as
\begin{align}\label{eq:bm}
\begin{split}
	\by\,|\,\bX,\bbeta,\sigma^2 &\,\sim \,N(\bX\bbeta,\sigma^2\mathbf{I}_n)  \\
	\bbeta\,|\,\sigma^2,\tau^2,\lambda_1^2,\ldots,\lambda_p^2 &\,\sim \,N(\boldsymbol{0},\sigma^2\tau^2\bD_{\blambda})\\
	 \bD_{\blambda}&=\mbox{diag}(\lambda_1^2,\ldots,\lambda_p^2)\\
	\lambda_j^2 &\,\sim \,\pi(\lambda_j^2)d\lambda_j^2,\quad  j=1,\ldots,p\\
	\tau &\,\sim \,C^+(0,1)\\
	\sigma^2 &\,\sim\, \sigma^{-2} d\sigma^2,
\end{split}
\end{align} 
where $\tau$ is the global shrinkage parameter, $\lambda_j$ are the local shrinkage parameters and $C^{+}(0,1)$ denotes the standard half-Cauchy distribution. Commonly used prior distributions for the local shrinkage parameters, $\lambda_j$, include
% for the Lasso, the horseshoe and the horseshoe+ are
%
\begin{equation}\label{eq:priors}
\begin{split}
	\lambda_j^2 \,\sim\, \mbox{Exp}(1),\quad j=1,\ldots,p& \quad \mbox{for the Lasso}, \\
	\lambda_j \,\sim \,C^+(0,1),\quad j=1,\ldots,p & \quad \mbox{for the horseshoe}, \\
	\lambda_j \,\sim \,C^+(0,\phi_j),\ \phi_j\,\sim\, C^+(0,1),\quad j=1,\ldots,p & \quad \mbox{for the horseshoe+},\\
\end{split}
\end{equation}
respectively, where ${\rm Exp}(1)$ denotes the standard exponential distribution. 
%We need some small amount of info here on the above methods.

%----------------------------------- Subsection ----------------------------------- %
\subsection{Selection of Variable Groups}

Although possessing many desirable theoretical, computational, and empirical properties, continuous shrinkage techniques so far have focused primarily on identifying important individual predictors. In many practical situations, it is however of great interest to select groups of variables, or find sparse combinations of grouped variables. 
%A common example is additive models in which each continuous predictor can be expanded into a linear combination of basis functions, and in which selection of important variables is equivalent to selection of entire groups of basis functions. Another important example arises 
This can be seen in additive models where each continuous predictor is expressed as a linear combination of basis functions such that selection of important predictors becomes equivalent to selection of groups (linear combinations) of the basis functions. This can also be seen in genome-wide association studies, in which the aim is to determine which single nucleotide polymorphisms are associated with a disease or trait. Since the single nucleotide polymorphisms can be conceptually grouped into different genes, and the genes can be grouped into different ``pathways'', it is of great interest to be able to select important groups of single nucleotide polymorphisms.

Under the assumption that a group structure is given, the aim is to find sparsity at both the group level, as well as within the groups themselves. An important group selection procedure is the group Lasso \citep{2006Yuanp4967} that solves the convex optimisation problem
\begin{align*}
	\hat{\bm{\beta}}(\kappa) = \argmin\limits_{\bbeta} \left\{ \|\by-\sum\limits_{g=1}^G\bX_g\bbeta_g\|_2^2 + \kappa \sum\limits_{g=1}^G \sqrt{s_g}\|\bbeta_g\|_2 \right\},
\end{align*}
where $\bX_g$ is the submatrix of $\bX$ containing those predictors in group $g$, $\bbeta_g$ is the coefficient vector for group $g$, $s_g$ is the number of predictors in group $g$, $\|\cdot\|_2$ is the Euclidean norm, and $\bbeta = (\bbeta_1^\prime,\ldots,\bbeta_G^\prime)^\prime$. A potential weakness of the group Lasso is that it only provides a sparse solution for a set of groups, and does not provide sparsity at an individual predictor level. To select individual predictors as well as groups of predictors, the sparse-group Lasso \citep{2010Friedmanp} was proposed
\begin{align*}
	\hat{\bbeta}(\kappa_1,\kappa_2) = \argmin\limits_{\bm{\beta}} \left\{ \|\by-\sum\limits_{g=1}^G\bX_g\bbeta_g\|_2^2 + \kappa_1 \sum\limits_{g=1}^G \|\bbeta_g\|_2 +\kappa_2\|\bbeta\|_1 \right\},
\end{align*} 
which introduces $\ell_1$-type penalisation at both variable and group levels; when $\kappa_2=0$, the sparse-group Lasso reduces to the group Lasso, and when $\kappa_1=0$ it reduces to the regular Lasso.
The Lasso-based approaches described above, as well as extensions of these methods, require the estimation of tuning parameters. Cross-validation \citep{1974Stonep111, 1995Kohavip11371145} methods are frequently used to meet this requirement; however, selection of tuning parameters by cross-validation may fail to achieve consistent variable selection \citep{1993Shaop486}. A particular weakness of all non-Bayesian sparse estimation procedures is the difficulty in obtaining confidence intervals for the resulting estimates \citep{2004Bayarrip5880}.

The Bayesian formulation of the sparse group regression model works by specifying a hierarchical or multi-level prior distribution over the regression parameters that incorporates prior beliefs about the group structure of the predictors. 
%A particular strength of the Bayesian approach is that all hyper-parameters can be estimated automatically from the Bayesian model without appealing to any secondary estimation technique.
The specific hierarchy ensures that the hyper-parameters can be automatically estimated by the inherent Bayesian procedure. Additionally, as the usual approach to analysing Bayesian models is through Monte-Carlo Markov chain (MCMC) methods, posterior samples can be used to easily obtain credible intervals for all parameters. 

Recently, \cite{2016Chenp665683} proposed a Bayesian sparse group selection model (BSGS), which is based on two-component mixture priors. Under the sparsity assumption, only a small number of groups are active in explaining the response variable, and only a small number of predictors are active within each of the active groups. The BSGS model consists of two layers of indicator variables: the group indicators which determine whether each group is active or not, and the individual indicators which determine whether each particular predictor is active or not. The BSGS uses a group-wise Gibbs sampler for the involved posterior sampling, and has demonstrated superior performance in selecting active groups, as well as identifying the active predictors within selected groups, in comparison to the sparse group Lasso. However, just as with the non-group based spike-and-slab methods, a problem with the two-component finite mixture model approach is the requirement for exploration of an exponentially growing model space, which can result in problems of computational intractability.

%----------------------------------- Subsection ----------------------------------- % 
\subsection{Decoupled shrinkage and selection methods}\label{sec:DSSM}

Bayesian regression estimators based on continuous shrinkage priors have their own limitation in that they do not automatically yield sparse solutions. Even though shrinkage priors such as the horseshoe do promote sparsity, the usual posterior mean or median estimates of the regression coefficients never equal zero almost surely. Sparse models, in which some of the regression coefficients equal zero exactly, are of great value since they are more interpretable, particularly when $p$ is large.

%Although sparse regression estimates have reduced predictive performance relative to dense estimates such as the posterior mean, they are useful as they produce a more interpretable predictive model, and improving the cost of computation and storage of the high-dimensional data.

%For this reason, variable selection techniques that provide a sparse estimate from a set of posterior samples have become an area of research interest, and a number of techniques have appeared in the literature; 
Therefore, it is highly desirable to develop some Bayesian variable selection techniques that are produce sparse solutions from non-zero posterior samples of regression coefficients. Progress has already been made in this regard; for example, \cite{2012Bondellp16101624} and \cite{2016Zhangp} use variable selection techniques based on credible regions to convert posterior samples to sparse estimates. 
%A recent, and flexible approach, called the decoupled shrinkage and selection (DSS) method \citep{2015Hannp435448}, utilises conventional penalised regression estimation to produce sparse estimates directly from a set of posterior samples. 
\cite{2015Hannp435448} use conventional penalised regression estimation to produce sparse estimates directly from the posterior samples, and name their method the decoupled shrinkage and selection (DSS) procedure.

 %a posterior variable selection summary named the decoupled shrinkage and selection (DSS) method has been proposed (\cite{2015Hannp435448}) for the Bayesian model selection approaches to find sparse estimator.
%In parameter estimation, a loss function, $L(\cdot)$, is normally used to quantify the difference between an estimated model and the true values of the data.

The posterior mean frequently exhibits excellent performance in terms of prediction \citep{2014vanderPasp25852618}; the key idea underlying the DSS is to determine a regression coefficient vector that is sparse while remaining close in some sense to its posterior mean estimate, and therefore retaining good predictive performance. The DSS procedure works by forming a new target vector
\begin{equation}\label{eq:ybar}
	\bar{\by} = \bX \bar\bbeta,
\end{equation}
where $\bar\bbeta =\E{}{\bbeta \,|\, \by}$ is the posterior mean of the regression coefficients; the vector $\bar\by$ is a smoothed version of the original data vector ${\bf y}$ that incorporates the effects of the shrinkage induced by the prior distribution on the coefficients $\bbeta$. From this, a DSS loss function $L(\cdot)$ can be defined
\begin{equation}\label{DSS}
	L(\bgamma) = \kappa \|\bgamma\|_0 +n^{-1} \| \bar\by - \bX\bgamma \|_2^2,
\end{equation}
where $\|\bgamma\|_0 = \sum_{j=1}^p \mathbbm{1}(\gamma_j \neq0)$ is the $\ell_0$-norm and $\kappa > 0$ is a penalisation parameter. This loss function is a combination of a goodness-of-fit term that measures the closeness of the regression coefficient $\bgamma$ to the posterior mean $\bar{\bbeta}$, plus a penalty term that measures the ``complexity'' of the model, and induces sparsity. The coefficient vector
\begin{equation}\label{DSS_betalambda}
	\bbeta_\kappa = \underset{\bgamma}{\operatorname{argmin}}\left\{  \kappa \|\bgamma\|_0 +n^{-1} \| \bX \bar\bbeta - \bX\bgamma \|_2^2\right\}.
\end{equation}
then represents a particular sparsification of $\bar\bbeta$, with the degree of sparsification determined by the choice of $\kappa$. For $\kappa = 0$, no sparsification occurs; for larger values of $\kappa$, the vector $\bbeta_\kappa$ becomes increasingly sparse.

The optimal solution is achieved through the trade-off between the number of variables in the model and the predictive performance in terms of the mean-squared prediction errors. If a variable $X_j$ has a strong association with the response variable, the posterior mean, $\bar\beta_j$, associated with this variable will tend to be large, and therefore, $X_j$ is more likely to be included in the final sparsified model. Conversely, if a variable $X_j$ has little or no effect on the response variable, it is more likely to be removed from the final model because $\bar\beta_j$ will be small. Therefore, the DSS algorithm builds connections between the Bayesian-type methods and the popular penalised likelihood approaches, and produces sparse estimates with good predictive performance.

%The best sparse estimate that approximates the mean of posterior samples is computed through the correlation of $X$, and thus, correlation in the data is also taken into account.

%----------------------------------- Subsection ----------------------------------- %
%\subsection{A heuristic of variable selection in DSS method}
For a given value of $\kappa$, the solution of (\ref{DSS_betalambda}), $\bbeta_\kappa$, is a potentially sparse coefficient vector. To use the DSS method, an obvious question is how to choose an appropriate value of $\kappa$. In the original DSS paper~\citep{2015Hannp435448}, two statistics were proposed to help visualise the predictive deterioration for varying values of $\kappa$: 
\begin{enumerate}
	\item the variation explained by a sparsified linear predictor
	\begin{equation}
		\label{varexp}
		\rho_\kappa^2 = \frac{n^{-1}\|\bX\bbeta \|^2 }{n^{-1}\|\bX\bbeta \|^2 +\sigma^2+n^{-1}\| \bX\bbeta-\bX\bbeta_\kappa \|^2 };
	\end{equation}
	
	\item the excess error of a sparsified linear predictor
	\begin{equation}
		\label{excerr}
		\psi_\kappa = \sqrt{n^{-1}\| \bX\bbeta-\bX\bbeta_\kappa \|^2  +\sigma^2} -\sigma.
	\end{equation}
\end{enumerate}
The smallest model for which the 90\% $\rho_\kappa^2$ credible interval contains $\E{}{ \rho_0^2 \,|\, \by}$, where $\rho_0^2$ is the variation explained by the posterior mean, is recommended as a good sparse approximation of $\bar\bbeta$. Although this approach produces a single model, it is heuristic in nature, and there is no compelling reason to select a 90\% credible interval, in place of say, a 95\% interval.

%----------------------------------- Subsection ----------------------------------- %
\subsection{Our contribution}

In this paper, we extend the Bayesian model with continuous global-local shrinkage priors to handle group structures that include overlapping and multilevel group structures; in particular, we extend the recent horseshoe and horseshoe+ priors for grouped predictor structures. After obtaining the posterior samples from the Bayesian models, we extend the decoupled shrinkage and selection method for grouped variables and apply the group non-negative garrotte to produce sparsified estimates. Generalised information criteria using posterior expected estimates of the degrees of freedom of the models in the group DSS model path are then used to select the final model, and perform group-wise selection.

\section{Grouped global-local shrinkage models}
\label{sec:BGM}

%In this section, we introduce overlapping and multi-level groups and the Bayesian model for the group structures.

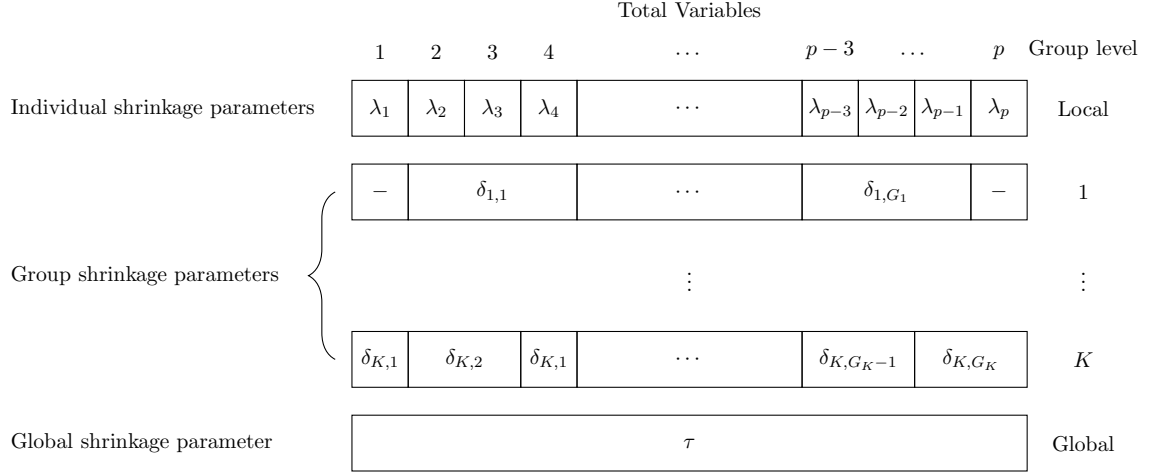
\begin{figure}
\centering
\begin{tikzpicture}[scale=0.74, every node/.style={scale=0.74}]
%\draw[step=1cm,green] (-2,-10) grid (16,2);
%---------------------Level 0--------------------------%
\node [above] at (8,2) {Total Variables};%\large \Large \LARGE \huge \Huge
% predictors 1 to p
\node [above] at (2.5, 1.25) {$1$};
\node [above] at (3.5, 1.25) {$2$};
\node [above] at (4.5, 1.25) {$3$};
\node [above] at (5.5, 1.25) {$4$};
\node [above] at (8, 1.25) {$\cdots$};
\node [above] at (10.5, 1.25) {$p-3$}; %\small \footnotesize \scriptsize \tiny
\node [above] at (12,1.25){$\cdots$};
%\node [above] at (11.5, 1.25) {\scriptsize $p-2$};
%\node [above] at (12.5, 1.25) {\scriptsize $p-1$};
\node [above] at (13.5, 1.25) {$p$};

\node [above] at (15, 1.25) {Group level};
%---------------------Level 1--------------------------%
% lambda's at level 1
\node [right] at (-4.2, 0.5) {Individual shrinkage parameters};

\draw (2,0) rectangle node{$\lambda_1$} +(1,1);
\draw (3,0) rectangle node{$\lambda_2$} +(1,1);
\draw (4,0) rectangle node{$\lambda_3$} +(1,1);
\draw (5,0) rectangle node{$\lambda_4$} +(1,1);
\draw (6,0) rectangle node{$\cdots$} +(4,1);
\draw (10,0) rectangle node{$\lambda_{p-3}$} +(1,1);
\draw (11,0) rectangle node{$\lambda_{p-2}$} +(1,1);
\draw (12,0) rectangle node{$\lambda_{p-1}$} +(1,1);
\draw (13,0) rectangle node{$\lambda_{p}$} +(1,1);

\node at (15, 0.5) {Local};
%---------------------Level 2--------------------------%
% delta's at level 2
\node [right] at (-4.2, -2.5) {Group shrinkage parameters};
\draw[decoration={brace,amplitude=10pt,raise=1},decorate]
  (1.75,-4) --  (1.75,-1);
  
\draw (2,-1.5) rectangle node{$ - $} +(1,1);
\draw (3,-1.5) rectangle node{$\delta_{1,1}$} +(3,1);
\draw (6,-1.5) rectangle node{$\cdots$} +(4,1);
\draw (10,-1.5) rectangle node{$\delta_{1,G_1}$} +(3,1);
\draw (13,-1.5) rectangle node{$-$} +(1,1);

\node at (15, -1) {1};
%---------------------Level 3--------------------------%
\node at (8, -2.5) {$\vdots$};
\node at (15, -2.5) {$\vdots$};
%---------------------Level 4--------------------------%
\draw (2,-4.5) rectangle node{$\delta_{K,1}$} +(1,1);
\draw (3,-4.5) rectangle node{$ \delta_{K,2} $} +(2,1);
\draw (5,-4.5) rectangle node{$\delta_{K,1}$} +(1,1);
\draw (6,-4.5) rectangle node{$\cdots$} +(4,1);
\draw (10,-4.5) rectangle node{$\delta_{K,G_K-1}$} +(2,1);
\draw (12,-4.5) rectangle node{$\delta_{K,G_K}$} +(2,1);

\node at (15, -4) {$K$};
%---------------------Level 5--------------------------%
% tau at level 5
\node [right] at (-4.2, -5.5) {Global shrinkage parameter};
\draw (2,-6) rectangle node{$\tau$} +(12,1);

\node at (15, -5.5) {Global};

\end{tikzpicture}
\caption{An illustration of possible group structures of $p$ variables with one level of individual variables, $K$ levels of grouped variables and one level of total variables.}\label{fig:groups}
\end{figure}

In regression, predictors with similar characteristics can often be collected into groups. For example, in bioinformatics, genetic variants used to predict the risk of a disease, can be formed into groups corresponding to particular genes. Generally, only a small number of genes are assumed to have important associations with the disease and we wish to select these important genes (i.e., perform sparse group selection). In addition, there exist many alternative groupings of genetic variants such as pathways, groups consisting of variants with similar function, etc. We propose to model such problems using a Bayesian multi-level hierarchy which can capture alternative schemes for grouping predictors (see Figure~\ref{fig:groups}).

We assume that each level of our multi-level hierarchy consists of non-overlapping groups. Two groups are said to be non-overlapping if they share no predictors in common. Assuming that groups within a level of our hierarchy are mutually exclusive greatly simplifies the sampling procedure. However, between different levels of the hierarchy, we allow the specification of overlapping groups which enables modelling of arbitrary grouping structures.

%Two groups are said to be overlapped if there is at least one element in common. In the hierarchical model, all individual levels consist of non-overlapping groups, but across different levels, groups can overlap across different levels. 

Consider a multi-level hierarchy consisting of $K+2$ levels. Within this multi-level hierarchy, we define a local level consisting of $p$ groups each containing a single predictor and a global level which consists of a single group containing all $p$ predictors. Levels $1$ through $K$ contain non-overlapping groupings of predictors determined by the application at hand. This approach generalises the global-local shrinkage hierarchy discussed in Section~\ref{sec:intro}. The local shrinkage parameters $(\lambda_1,\ldots,\lambda_p)$ that control the amount of shrinkage for individual predictors are located at the first level of the hierarchy. At the last level of the hierarchy is the global shrinkage parameter $\tau$ which determines the overall level of shrinkage for all predictors. To perform estimation and selection for grouped variables within a global-local shrinkage hierarchy, we introduce group shrinkage parameters associated with each group of predictors. We denote the group shrinkage parameter for the $g$-th group at the $k$-th level of the hierarchy by $\delta_{k,g}$. These group shrinkage parameters induce an additional level of shrinkage on all predictors in a group.

Figure~\ref{fig:groups} shows an example of a possible multilevel group structure. Suppose there are $p$ variables and $K+2$ levels of specified groups of predictors. At the first group level, each of the $p$ predictors belongs to a separate group of size one. In the last level, all predictors are grouped together into one group of size $p$. Figure~\ref{fig:groups} clearly demonstrates the flexibility of our proposed multi-level approach. There is no requirement that each predictor be assigned to a group at any of the $K$ levels outside the local and the global level; for example, in level $1$ variables $1$ and $p$ are not assigned to a group. In addition, the groups are not required to be contiguous and can contain any combination of the $p$ predictors; for example, group $1$ at level $K$ contains both predictor $1$ and predictor $4$.

%In between the first and last levels, are different group structures where variables are grouped with different combinations. 

%For the Lasso, horseshoe and horseshoe+ models, the shrinkage parameters $(\lambda_1,\ldots,\lambda_p)$ at the first level are a series of group shrinkers for a single variable. The global shrinker, $\tau$, at the last level can be treated as a single group shrinker for all variables in the model. The group shrinkage parameters $\delta_{k,g}$ penalise all variables in the same group simultaneously.

%-------------------- Diagram: group structures --------------------%
%-------------------- End of Diagram: group structures --------------------%

%-------------------------------------------------------------------------
\subsection{Bayesian grouped global-local shrinkage hierarchical models}
\label{sec:BGGLSHM}
Formally, let $G(k,j) \in \{ 0,\ldots,G_k\}$ denote the group containing predictor $j$ at level $k$ of our hierarchy where $G_k \geq 1$ is the number of groups defined at level $k$. If predictor $j$ does not belong to any group at level $k$ we set $G(k,j)=0$. The hierarchical representation of our proposed Bayesian grouped global-local shrinkage regression model can then be written as
\begin{align}
	\by \,| \,\bX,\bbeta,\sigma^2 & \,\sim  \,\mathcal{N} (\bX\bbeta,\sigma^2\mathbf{I}_n) \nonumber \\
	\bbeta \,| \,\sigma^2,\tau^2,\lambda_j,\delta_{k,g} & \,\sim  \,\mathcal{N} (\boldsymbol{0},\sigma^2\tau^2\bD_{\blambda}\bD_{\bdelta_1} \cdots \bD_{\bdelta_K}),   \nonumber\\
	\lambda_j  & \,\sim \,  \pi(\lambda_j)d\lambda_j, &\quad j&=1,\ldots,p  \nonumber\\
	\delta_{k, g} & \,\sim \, \pi(\delta_{k, g})d \delta_{k, g}, &\quad g&=1,\ldots,G_k,\ k=1,\ldots,K \nonumber\\
	\tau & \,\sim  \,C^+(0,1)\nonumber \\
	\sigma^2 & \,\sim \, \sigma^{-2} d\sigma^2 \label{BGLSRM}
\intertext{where}
	\bD_{\blambda} & \,=\, \mbox{diag}(\lambda_1^2,\ldots,\lambda_p^2)\nonumber\\
	\bD_{\bdelta_k} & \, = \, \mbox{diag}(\Omega_{k,1},\ldots,\Omega_{k,p}), & \quad k & =1,\ldots,K \label{eq:bDmatrix}\\ 
	\Omega_{k,j} & \, =\,  \mathbbm{1} \{\omega_{k,j}=0 \} + \omega_{k,j}\cdot\mathbbm{1} \{\omega_{k,j}\neq0 \},& \quad j & =1,\ldots,p,\  k=1,\ldots,K \nonumber\\ 
	\omega_{k,j} & \, =\,  \sum_{g=1}^{G_k} \delta_{k,g}^2 \cdot \mathbbm{1} \{ G(k,j) = g \},& \quad j & =1,\ldots,p,\  k=1,\ldots,K \nonumber 
\end{align}
Here, $\blambda=(\lambda_1,\ldots,\lambda_p)$ are the local shrinkage parameters associated with the individual predictors and $\bdelta_1=(\delta_{1,1},\ldots,\delta_{1,G_1}),\ldots,\bdelta_K=(\delta_{K,1},\ldots,\delta_{K,G_K})$ are the group shrinkage parameters.

The global shrinkage parameter, $\tau$, controls the overall degree of shrinkage of the coefficients while the local shrinkage parameters, $\lambda_j$, determine the shrinkage for each individual coefficient. The grouped shrinkage parameters $\delta_{k,j}$ provide additional shrinkage for predictors within groups. In the special case where $K=0$ (i.e., there are only a local and a global levels), the model reduces to the usual Bayesian global-local shrinkage regression model (\ref{eq:bm}).

%This model offers more flexibility than the regular Bayesian model (\ref{eq:bm}) by shrinkage groups of predictors individual, and also allows shares information between predictors to shrink at a group level.

This model offers more flexibility than the regular Bayesian model (\ref{eq:bm}) by shrinking individual predictors separately, and allowing shrinkage for groups of predictors simultaneously. The ability to specify complex group structures using multiple levels further increases the flexibility of the proposed model. In many real-world applications, each predictor may belong to more than one group. For example, in statistical genomics, predictors corresponding to single nucleotide polymorphisms may belong to different genes, and a collection of genes may form a biological pathway. Additionally, pathways may share genes and genes may share single nucleotide polymorphisms. By utilising an appropriate number of grouping levels, the complexity of this situation is easily handled by our proposed hierarchical model.

After the model is specified, posterior samples of $\bbeta$, $\sigma^2$, $\tau$, $\lambda_j$, $\delta_{k, g}$ can be obtained from the full conditional distributions through the implementation of a Gibbs sampler, which cyclically updates one parameter at a time given the current values of all the other parameters and hyper-parameters. To sample the global hyper-parameter $\tau$, we use the sampler proposed by \cite{2016Enesp179182} which decomposes 
%
%The only remaining concern is that the conditional posterior distribution for the hyper-parameter, $\tau$, is difficult to sample from using the Gibbs sampler due to the non-standard form caused by the half-Cauchy distribution. This issue can be addressed by proposing a simple sampler (\cite{2016Enesp179182}) that enables a straightforward sampling of the full conditional posterior distributions by decomposing
%
\begin{equation}\label{halfcauchy}
\tau \,\sim\, C^+(0,a)
\end{equation}
into 
\begin{equation}
\label{cauchydecomp}
	\tau^2 \,| \nu \,  \sim \mathcal{IG}\left( \frac{1}{2},\frac{1}{\nu} \right) \ \mbox{and} \  \nu \, \sim \, \mathcal{IG} \left( \frac{1}{2}, \frac{1}{a^2} \right),
\end{equation}
where $\mathcal{IG}$ denotes the inverse gamma distribution.

%Irrespective of the choice of prior distributions for the local shrinkage parameters $\lambda_j$ and group shrinkage parameters $\delta_{k,g}$, 
The posterior distributions for $\bbeta$, $\sigma^2$ and $\tau$ do not depend on the choice of prior distributions for the local shrinkage parameters $\lambda_j$ and group shrinkage parameters $\delta_{k,g}$. Therefore, the full conditional distributions of $\bbeta$, $\sigma^2$ and $\tau$ are the same for the Bayesian group Lasso, the Bayesian group horseshoe and the Bayesian group horseshoe+ models. The full conditional distribution of $\bbeta$ is
\begin{equation*}
\begin{split}
	&\bbeta \, | \, \cdot \,  \sim  \, \mathcal{N} \left(\bA^{-1}\bX^T\by,\sigma^2\bA^{-1}\right)\\
	& \bA = \bX^T \bX+(\tau^2 \bD_{\blambda}\bD_{\bdelta_1} \cdots \bD_{\bdelta_K} )^{-1}.
\end{split}
\end{equation*}

The full conditional distribution of $\sigma^2$ is
\begin{equation*}
	\sigma^2 \, | \, \cdot  \, \sim \,  \mathcal{IG}\left(\frac{n-1+p}{2},\frac{(\by-\bX\bbeta)^T(\by-\bX\bbeta)+\bbeta^T(\tau^2\bD_{\blambda}\bD_{\bdelta_1} \cdots \bD_{\bdelta_K})^{-1}\bbeta}{2}\right).
\end{equation*}

The full conditional distributions of $\tau$ and $\nu$ are
\begin{equation*}
\begin{split}
	\tau^2 \, |\cdot& \,  \sim \,  \mathcal{IG}\left(\frac{p+1}{2},\frac{\bbeta^T(\bD_{\blambda}\bD_{\bdelta_1} \cdots \bD_{\bdelta_K})^{-1}\bbeta}{2\sigma^2}+\frac{1}{\nu}\right)\\
	\nu  \, | \, \cdot & \, \sim \mathcal{IG}\left(1,\frac{1}{\tau^2}+1\right),
\end{split}
\end{equation*}
where sampling from $\mathcal{IG}\left(1,1/\tau^2+1\right)$ is equivalent to sampling from an exponential distribution. The full conditional distributions of $\lambda_j$ and $\delta_{k, g}$ depend on the particular choice of the prior distribution and are discussed in the next section.
%, such as the Laplace, horseshoe and horseshoe+.

%----------------------------------- Subsection ----------------------------------- %
\subsection{Bayesian group Lasso model}
\cite{2008Parkp681686} proposed the Bayesian Lasso in which each regression coefficient follows an independent double-exponential (Laplace) distribution. \cite{2010Kyungp369412} extended this idea to the Bayesian group Lasso, which we now further extend to our multilevel group hierarchy. The Laplace prior of the Lasso model can be expressed as a Gaussian variance mixture distribution where the mixing density is an exponential distribution. This implies that the Bayesian Lasso can be written as a Bayesian global-local shrinkage hierarchy where the prior distributions for the local shrinkage parameters $\lambda_j$ are
\begin{equation}
	\label{eq:lassopriors}
	\lambda_j^2 \,  \sim \,  \mbox{Exp}(1),\  j=1,\ldots,p.
\end{equation}
Let $\bdelta$ denote the collection of $\delta_{k,g}$, the full conditional distribution of $\lambda_j^{-2}$, $j=1,\ldots,p$ is then
\begin{equation*}
\lambda_j^{-2} \, | \, \bbeta,\sigma^2,\tau^2,\bdelta \, \sim  \, \mbox{InvGaussian}\left(  \alpha_j ^{\frac{1}{2}},2\right),
\end{equation*}
where 
\begin{equation*}
%\alpha_j = \frac{2\sigma^2\tau^2 \prod\limits_{k=1}^K\delta^2_{k, j}}{\beta_j^2}.
	\alpha_j = \frac{2\sigma^2\tau^2 [\bD]_{j,j} }{\beta_j^2}\quad \mbox{and}\quad \bD = \bD_{\bdelta_1}\cdots\bD_{\bdelta_K}.
\end{equation*}
Here $[\bD]_{j,j}$ is the entry in the $j$th row and $j$th column of the matrix $\bD$.
%\textsuperscript{th}
%
Now we choose the prior distribution for the group shrinkage parameter $\delta_{k,g}^2$ associated with group $g$ at level $k$ to be
\begin{equation}
	\delta_{k,g}^2  \, \sim  \, \mbox{Exp}(1),\  g=1,\ldots,G_k,\ k=1,\ldots,K,
\end{equation}
where $G_k$ is the number of predictor groups at level $k$ of our hierarchy. The full conditional distribution of $\delta_{k,g}^{-2}$ is then a generalised inverse Gaussian (GIG) distribution of the form
\begin{equation*}
	\delta_{k, g}^{-2} \,  | \, \bbeta,\sigma^2,\tau^2,\blambda  \, \sim \,  \mbox{GIG}\left( \alpha_{k,g},2 , \frac{s_{k,g}}{2}-1 \right),
\end{equation*}
where 
\begin{equation*}
%\alpha_k =\frac{2\sigma^2\tau^2}{\bbeta_{g_k}'\left(\bD_{\blambda_{g_k}} \prod\limits_{j=1,j\neq k}^K \bD_{\delta_j}\right)  \bbeta_{g_k}}.
	\alpha_{k,g} =\frac{1}{\sigma^2\tau^2} \sum_{i \in l(k,g)} \frac{ \beta_i^2 }{\lambda_i^2 [\bD_{-k}]_{i,i}},
\end{equation*}
$s_{k,g}$ is the number of predictors in group $g$ at level $k$ and $l(k,g) = \{j \in \{1,\ldots,p \} : G(k,j) = g \}$ is the set of predictors at level $k$ that belongs to group $g$. Here 
\begin{equation}
\label{eq:D-k}
	\bD_{-k} =\prod_{i=1,i\neq k}^K \bD_{\bdelta_i}
\end{equation}
where $\bD_{\bdelta_i}$ is given by~\eqref{eq:bDmatrix}.

%----------------------------------- Subsection ----------------------------------- %
\subsection{Bayesian group horseshoe model}
\label{sec:BGHSM}
The horseshoe prior has been demonstrated to have good performance in high dimensional and sparse regression problems, leaving strong signals untouched and aggressively shrinking noise variables \citep{2009Carvalhop7380}. The prior distribution for the local shrinkage parameters, $\lambda_j$, in the horseshoe model is
\begin{equation}
	\label{eq:priors3}
	\begin{split}
	\lambda_j  \, \sim \,  C^+(0,1),\ j=1,\ldots,p, \\
	\end{split} 
\end{equation}
where $C^+(0,1)$ is the standard half-Cauchy distribution.
By introducing auxiliary variables in similar fashion to (\ref{halfcauchy}) and (\ref{cauchydecomp}), the standard half-Cauchy distribution can be expressed as
\begin{equation*}
	\lambda_j^2 \, | \, c_j  \, \sim  \, \mathcal{IG}\left( \frac{1}{2},\frac{1}{c_j} \right), \  c_j \, \sim \, \mathcal{IG} \left( \frac{1}{2},1 \right).
\end{equation*}
The full conditional distributions for $\lambda_1^2,\ldots,\lambda_p^2$ and the auxilary variables $c_1, \ldots, c_p$ are
\begin{equation*}
\begin{split}
	\lambda_j^2 \,  | \, \bbeta,\sigma^2,\tau^2,\bdelta,c_j  &\, \sim  \, \mathcal{IG}\left(1,\frac{1}{\alpha_j} +\frac{1}{c_j} \right),\quad \alpha_j =\frac{2\sigma^2\tau^2 [\bD]_{j,j}}{\beta_j^2}\\
	c_j \, | \, \lambda_j^2 & \, \sim  \, \mathcal{IG} \left( 1,\frac{1}{\lambda_j^2}+1\right).
\end{split}
\end{equation*}
The group shrinkage parameters $\delta_{k,g}^2$ in the group horseshoe also follow the standard half-Cauchy distribution~\citep{2016Xup229240}
\begin{equation}
\delta_{k,g} \,  \sim  \, C^+(0,1),\  g=1,\ldots,G_k,\ k=1,\ldots,K \, .
\end{equation}
Using the half-Cauchy decomposition from \eqref{halfcauchy} and \eqref{cauchydecomp}, these prior distributions can also be written as
\begin{equation}
	\delta_{k,g}^2 \, | \,  t_{k,g} \sim \mathcal{IG}\left( \frac{1}{2},\frac{1}{t_{k,g} } \right),\  t_{k,g} \,  \sim \, \mathcal{IG} \left( \frac{1}{2},1 \right).
\end{equation}
The full conditional distributions for $\delta_{k,g}^2$ and $t_{k,g}$ are then
\begin{equation*}
\begin{split}
	\delta_{k, g}^2 \, | \, \bbeta,\sigma^2,\tau^2,\blambda, t_{k, g} &\,  \sim \,  \mathcal{IG}\left(\frac{s_{k,g}+1}{2}, \frac{1}{2\sigma^2\tau^2} \sum_{i\in l(k,g)} \frac{\beta_i^2}{\lambda_i^2 [\bD_{-k}]_{i,i}} + \frac{1}{t_{k, g}}\right) \\
t_{k,g} \, | \, \delta_{k, g}^2 &\,  \sim \,  \mathcal{IG}\left(1,\frac{1}{\delta_{k, g}^2}+1\right) ,
\end{split}
\end{equation*}
where $s_{k,g}$ is the number of predictors in group $g$ at level $k$, $l(k,g) = \{j \in \{ 1,\ldots,p \} : G(k,j) = g \}$ is the set of predictors at level $k$ that belongs to group $g$ and $\bD_{-k}$ is given by~\eqref{eq:D-k}.

%
%
%----------------------------------- Subsection ----------------------------------- %
\subsection{Bayesian group horseshoe+ model}
%
%
%To handle the ultra-sparse problem, the horseshoe prior can be extended to incorporate an extra level of hyper-parameters and the horseshoe+ prior is proposed. 
The horseshoe+ prior has a heavier tail than the regular horseshoe prior, resulting in more aggressive shrinkage of weak signals and producing sparser estimates than the regular horseshoe prior. The prior distribution of the local shrinkage parameters, $\lambda_j$, for the horseshoe+ model is
\begin{equation}\label{eq:priors4}
\begin{split}
	\lambda_j \, | \,\phi_j \, \sim \,  C^+(0,\phi_j),\ \phi_j \, \sim \,  C^+(0,1),\ j=1,\ldots,p. \\
\end{split}
\end{equation}
By using the decomposition in (\ref{halfcauchy}) and (\ref{cauchydecomp}), the above prior can be written as
\begin{equation*}
	\lambda_j^2 \, | \, c_j  \, \sim  \, \mathcal{IG}\left( \frac{1}{2},\frac{1}{c_j} \right),\ c_j \, | \, \phi_j^2 \, \sim \, \mathcal{IG} \left( \frac{1}{2},\frac{1}{\phi_j^2} \right),\ \phi_j^2 \, | \, \eta_j  \, \sim  \, \mathcal{IG}\left(\frac{1}{2},\frac{1}{\eta_j} \right),\ \eta_j \, \sim \, \mathcal{IG}\left(\frac{1}{2},1 \right).
\end{equation*}
The full conditional distributions for $\lambda_j^2$ and associated auxilary variables are
\begin{equation*}
\begin{split}
	\lambda_j^2 \,  | \, \bbeta,\sigma^2,\tau^2,\bdelta,c_j&  \, \sim  \, \mathcal{IG}\left(1,\frac{\beta_j^2}{\alpha_j} +\frac{1}{c_j} \right),\quad \alpha_j=2\sigma^2\tau^2 [\bD]_{j,j}\\
	c_j \, | \, \lambda_j^2, \phi_j^2 & \, \sim  \, \mathcal{IG} \left( 1,\frac{1}{\lambda_j^2}+\frac{1}{\phi_j^2} \right)\\
	\phi_j^2 \,  |  \, c_j, \eta_j & \, \sim \,  \mathcal{IG}   \left( 1, \frac{1}{c_j}+\frac{1}{\eta_j} \right)\\
	\eta_j \, | \, \phi_j^2  &\, \sim  \, \mathcal{IG} \left( 1,\frac{1}{\phi_j^2}+1  \right).
\end{split}
\end{equation*}
The horseshoe+ prior distribution for the group shrinkage parameters $\delta_{k,g}^2$ is
\begin{equation*}
	\delta_{k,g}\, | \,t_{k,g}  \, \sim \,  C^+(0, t_{k,g}),\ t_{k,g} \, \sim \,  C^+(0,1).
\end{equation*}
By using the decomposition, the above prior can be written as
\begin{equation*}
\begin{split}
	\delta_{k,g} ^2 \, | \, \xi_{k,g}   &\, \sim  \, \mathcal{IG}\left( \frac{1}{2},\frac{1}{\xi_{k,g} } \right),\ \xi_{k,g} \,  | \, t_{k,g} ^2 \, \sim \, \mathcal{IG} \left( \frac{1}{2},\frac{1}{t_{k,g} ^2} \right)\\
	t_{k,g} ^2 \, | \, \psi_{k,g}   &\, \sim  \, \mathcal{IG}\left(\frac{1}{2},\frac{1}{\psi_{k,g} } \right),\ \psi_{k,g} \,  \sim \, \mathcal{IG}\left(\frac{1}{2},1 \right).
\end{split}
\end{equation*}
Using the decomposition \eqref{halfcauchy} and \eqref{cauchydecomp}, the full conditional distributions of $\delta_{k,g}^2$ and the associated auxiliary variables are 
\begin{equation*}
\begin{split}
	\delta_{k, g}^2 \, | \, \bbeta,\sigma^2,\tau^2,\blambda, t_{k, g} &\,  \sim \,  \mathcal{IG}\left(\frac{s_{k,g}+1}{2}, \frac{1}{2\sigma^2\tau^2} \sum_{i\in l(k,g)} \frac{\beta_i^2}{\lambda_i^2 [\bD_{-k}]_{i,i}} + \frac{1}{\xi_{k, g}}\right) \\
	\xi_{k,g} \,  | \,  \delta_{k, g}^2,t_{k,g}^2 & \, \sim  \, \mathcal{IG}\left(1,\frac{1}{\delta_{k, g}^2}+\frac{1}{t_{k,g}^2} \right)\\
	t_{k,g}^2 \, | \,  \xi_{k,g},\psi_{k,g}& \,  \sim \,  \mathcal{IG} \left( 1,\frac{1}{\xi_{k,g}} + \frac{1}{\psi_{k,g}} \right) \\
	\psi_{k,g} \, | \, t_{k,g}^2 & \, \sim \,  \mathcal{IG} \left( 1, \frac{1}{t_{k,g}^2} +1 \right).
\end{split}
\end{equation*}
where $s_{k,g}$ is the number of predictors in group $g$ at level $k$, $l(k,g) = \{j \in \{1,\ldots,p\} : G(k,j) = g \}$ is the set of predictors at level $k$ that belongs to group $g$, and $\bD_{-k}$ is given by~\eqref{eq:D-k}.

%----------------------------------- Subsection ----------------------------------- %
%\subsection{Selecting various prior distributions}
%The hierarchical representations in (\ref{BGLSRM}) offer great flexibility for different modeling approaches. Different prior distributions at various levels can be chosen for different models. For example, the horseshoe prior can be chosen for the local shrinkage parameters, $\blambda$, for individuals and the Lasso prior can be chosen for the group shrinkage parameters, $\bdelta$. If there are no local shrinkage parameters for individuals and the prior distribution of one level of group shrinkage parameters is the exponential distribution, then the model becomes the regular Bayesian grouped Lasso model.

%================================================%
%===================== Section =====================%
\section{Decoupled shrinkage and selection for grouped variables}
Recall that in the DSS procedure, the sparsification of $\bar\bbeta$ involves solving a penalised regression problem using the smoothed data vector $\bar{\by} = \bX \bar\bbeta$:
\begin{equation}\label{DSS_beta}
	\bbeta_\kappa = \underset{\bbeta}{\operatorname{argmin}}\left\{  n^{-1} \| \bar{\by} - \bX\bbeta \|_2^2 + \kappa \|\bbeta\|_0 \right\}.
\end{equation}
There are two potential limitations of the original DSS procedure: (i)~it does not generalise to the problem of selecting groups of variables, and (ii)~it does not provide an objective and general procedure for selecting which of the models in the DSS solution path should be used as the final, sparsified estimate. In this section we propose an extension of the DSS procedure, referred to as group DSS, which extends the original DSS proposal to accommodate selection of groups, and discuss how to adapt the standard information criteria based approaches to model selection to the DSS framework.

%, it requires a further step to select the regularisation parameter, $\kappa$, and produce a final solution $\bbeta_\kappa$. 
%
%
%
%----------------------------------- Subsection ----------------------------------- %
\subsection{Group decoupled shrinkage and selection methods}\label{sec:GDSS}

The original DSS method produces sparse estimates at the level of individual variables rather than groups. Consider the multi-level group hierarchy discussed in Section~\ref{sec:BGM}. If we restrict interest to sparse estimation at a single level of the hierarchy, the ideas underlying DSS can easily be extended to deal with grouped variables, as none of the groups at a given level overlap. For the purposes of the group DSS procedure, variables that do not belong to any group are considered to form their own singleton groups. 
%
%
%Let $G$ denote the number of groups at the level of interest, including those groups that contain only a single variable, so that
% 
%\[
%	\bm{\mu} =  \sum_{g=1}^G {\bf X}_g \bbeta_g ,
%\]
%%
%where $\bX_g$ is the submatrix of $\bX$ containing those predictors in group $g$ and $\bm{\beta}_g$ are the regression coefficients associated with group $g$.
%
To select sparse estimates for grouped variables, we propose to use the solution to the following ideal group DSS optimisation problem:
%\begin{equation}\label{DSSg}
%	L(\bgamma) = \lambda \sum\limits_{g=1}^G \mathbbm{1}( \|\bgamma_g\|_0\neq 0) +n^{-1} \| \bX \bar\bbeta - \sum\limits_{g=1}^G \bX_g\bgamma_g \|_2^2,
%\end{equation}
%
%where $G$ is the total number of groups. The optimal solution can be expressed as:
% $n^{-1} \| \bX \bar\bbeta - \sum\limits_{g=1}^G \bX_g\bbeta_g \|_2^2$
%
\begin{equation}
	\label{betaDSSg}
	{\bbeta_\kappa}=\underset{\bbeta_1,\ldots,\bbeta_G}{\operatorname{argmin}}\left\{  n^{-1} \| \bX \bar\bbeta - \sum\limits_{g=1}^G \bX_g\bbeta_g \|_2^2   + \kappa\sum\limits_{g=1}^G s_g \mathbbm{1}( \|\bbeta_g\|_0\neq 0) \right\},
\end{equation}
where $\bX_g$ is the submatrix of $\bX$ containing all predictors in group $g$, $\bm{\beta}_g$ are the regression coefficients associated with group $g$, $s_g$ is the number of predictors in group $g$ and $G$ is the total number of groups at the level of the hierarchy we are interested in.

However, one potential problem in solving the optimisation problem (\ref{betaDSSg}) is that the counting penalty $\|\bbeta_g\|_0$ results in a potentially intractable optimisation problem when the number of groups is moderate to large. A standard approach to address this issue is to approximate the $\ell_0$ penalty with an $\ell_1$ penalty; for example, in the case of the original DSS algorithm (\ref{DSS_betalambda}), \cite{2015Hannp435448} proposed the following surrogate $\ell_1$ optimisation problem 
\begin{equation*}
\bbeta_\kappa = \underset{\bgamma}{\operatorname{argmin}}\left\{ \kappa \sum_{j=1}^p \frac{|\gamma_j|}{|w_j|}  + n^{-1} \| \bX \bar\bbeta - \bX\bgamma \|_2^2 \right\},
\end{equation*} 
where $w_j=\bar\beta_j$ is the posterior mean for coefficient $j$. This approach is analogous to the adaptive Lasso~\citep{2006Zoup14181429} in which $w_j$ is replaced by the least-squares estimate $\hat{\beta}_j$, and can be easily solved by a rescaling of the design matrix $\bX$.

A similar approach may be used to extend DSS to grouped variables. For example, we could replace the $\ell_0$-norm in (\ref{betaDSSg}) by the grouped Lasso penalty~\citep{2006Yuanp4967}
%In terms of the grouped model, the optimal solution in (\ref{betaDSSg}) is related to the group Lasso estimate proposed by :
%
\begin{equation*}
\frac{1}{2}\| \by - \sum\limits_{g=1}^G \bX_g\bbeta_g \|_2^2 + \kappa\sum\limits_{g=1}^G \| \bbeta_g \|_{\bK_g},
 \end{equation*}
where $ \| \bbeta_g \|_{\bK_g} = (\bbeta_g^T \bK_g \bbeta_g)^{1/2}$ and $\bK_g$ are some symmetric positive definite matrices. However, a difficulty in using the group Lasso is that the corresponding solution path is no longer piecewise linear and is potentially slow to compute.

In this paper, we use the group non-negative garrotte (NNG) in place of the group Lasso. We adopted this procedure as it is relatively simple to use, and its formulation is similar in nature to the adaptive Lasso. Furthermore, the group NNG reduces the dimensionality of the solution path to the number of groups. An approximate solution to (\ref{betaDSSg}) can be achieved through the group non-negative garrotte by solving
\begin{equation}
	\label{DSSnng}
	\bd_\kappa =\underset{\bd}{\operatorname{argmin}} \left( \frac{1}{2} \|\bar\by-\bZ \bd\|^2 +\kappa \sum\limits_{g=1}^G s_g d_g \right),\quad\mbox{subject to } d_g\geq0,\forall g ,
\end{equation}
where $\bar\by = \bX\bar\bbeta$, $\bZ=(\bZ_1,\ldots,\bZ_G)$, $\bZ_g = \bX_g\bar\bbeta_g$, $s_g$ is the size of group $g$ and ${\bf d} = (d_1,\ldots,d_G)$ is a vector of group shrinkage coefficients. The algorithm for computing the non-negative garotte solution path is similar to the group least angle regression algorithm, and the solution to the group DSS for a given $\kappa$ can be computed as $\bbeta_\kappa=(\bar\bbeta_1 [\bd_\kappa]_1,\ldots,\bar\bbeta_G [\bd_\kappa]_G)$.

The group DSS described above only works on a single level of group structure. However, it can be easily extended to multilevel group structures; for example, after obtaining posterior samples, group DSS can be run separately at each level to select groups of variables for every single level.

%----------------------------------- Subsection -------------------------------------- %
\subsection{Information criteria for group DSS}\label{sec:ICGDSS}
The group DSS procedure produces a number of candidate sparse models by varying the regularisation parameter $\kappa$ in~\eqref{DSSnng}, but does not specify which model should be chosen. To assist in selecting a sparse model, the original DSS procedure recommended a heuristic selection criterion based on credible intervals of two statistics: (i)~the variation explained (\ref{varexp}), and (ii)~the excess error (\ref{excerr}). Although both of these heuristics are easily adapted to grouped variables, the corresponding selection criteria still require the arbitrary choice of a credible interval size. We now propose an alternative approach to selecting $\kappa$ by adapting the standard information criteria approach commonly used in the model selection literature.

A standard approach for selecting a sparse linear model from a set of candidate models is through the use of generalised information criteria (GIC). The usual formulation of the generalised information criteria consists of a likelihood function that measures the goodness-of-fit of a model under consideration, and a model complexity penalty term that is a function of the degrees of freedom of the model; namely
\begin{equation}
	\label{GIC}
	\mbox{GIC}(\bbeta,\sigma^2) =  L(\bbeta,\sigma^2) + \alpha(k,n,\sigma^2)
\end{equation} 
%
%where $\bbeta_\kappa$ are the model coefficients, $\sigma^2$ is the unknown noise variance, $L(\cdot)$ is the negative log-likelihood function, 
in the context of linear regression model selection, where $L(\cdot)$ is
the negative log-likelihood function, $\bbeta$ are the model coefficients, $\sigma^2$ is the unknown noise variance, $\alpha(\cdot)$ is the penalty function, $n$ is the sample size and $k = ||\bbeta||_{\infty}$ is the degrees of freedom of the model. We then select the model that attains the minimum GIC value over the set of candidate models we are considering.

There exists a wide range of information criteria in the statistical model selection literature; one of the most well-known information criterion is the Bayesian information criterion (BIC)~\citep{1978Schwarzp461464}, which is given by
\begin{equation}
	\mbox{BIC}(\bbeta,\sigma^2) = L(\bbeta,\sigma^2) + \frac{k}{2}\ln(n),
\end{equation} 
where $n$ is the number of observations. The BIC is derived as an asymptotic approximation of the marginal probability of the data. Alternatives to BIC, that utilise more accurate approximations at the cost of stronger assumptions, include information theoretic minimum message length (MML)~\citep{2005WallaceSIIMML} and minimum description length~\citep{1983Rissanenp416431} principles. A recent MML criterion for linear-Gaussian regression models~\citep{2009Schmidtp312321} is
\begin{equation}\label{MMLu}
	\mbox{MML}_u(\bbeta,\sigma^2)  = L(\bbeta,\sigma^2) + \left(\frac{k+1}{2}\right)\ln \left(\frac{\| \by \|^2}{2\sigma^2}  \right ) - \Gamma \left(\frac{k+3}{2}\right) + \frac{1}{2}\ln \left(k+1\right),
\end{equation}
%
%The MML$_u$ criterion is statistically consistent under the fixed $p$, increasing $n$ scenario under standard assumptions. 
which has been shown to perform well in a wide range of settings~\citep{2011Giurcaneanup16711692}. In contrast to BIC, the MML$_u$ penalty term is a function of the signal-to-noise ratio of the fitted model.

%; the second term in (\ref{MMLu}) can be interpreted as the logarithm of $n$ times the signal-to-noise ratio plus one of the fitted model, so that the MML$_u$ penalty is also determined by how well the model fits the data.

Another widely used information criterion is the Akaike information criterion (AIC)~\citep{1973Akaikep267281}
\begin{equation}
	\mbox{AIC}(\bbeta,\sigma^2) = L(\bbeta, \sigma^2) + k.	
\end{equation}
The AIC penalises the number of parameters less strongly than the BIC, and in the case when the sample size $n$ is small, or the degrees of freedom $k$ is large relative to $n$, the AIC tends to select models with too many parameters, and therefore has an increased probability of overfitting. To address this issue, \cite{1989Hurvichp297307} proposed the corrected AIC 
\begin{equation}\label{eq:aicc}
	\mbox{AIC}_c(\bbeta,\sigma^2) = L(\bbeta,\sigma^2) +\frac{kn}{n-k-1}.
\end{equation}
Details regarding derivations and properties for AIC, BIC and related criteria can be found in~\cite{1998McQuarriep}.

%An even greater number of model selection methods exist in the specific case of linear regression with Gaussian disturbances, due to both the importance and mathematical tractability of this particular model. 

%-----------------------------------------------------------
%----------------------------------- Subsubsection ----------------------------------- %
\subsubsection{Information criteria for DSS}

Selecting a model using the GIC (\ref{GIC}) approach requires the existence of a negative log-likelihood function which measures how well each candidate model fits the data. In the DSS framework, we estimate each model for a given $\kappa$ by a penalised regression onto a smoothed data set $\bar{\by} = \bX \bar{\bbeta}$. Then model selection is equivalent to choosing the $\kappa$ value that minimises a GIC type criterion. By construction, the unpenalised model in the regression path corresponding to $\kappa = 0$ will fit the data $\bar{\by}$ perfectly with no errors, resulting in an infinite log-likelihood. Therefore, a direct application of the GIC is not possible without some modification.

Consider instead the Kullback--Leibler (KL)~\citep{1951Kullbackp7986} divergence from a probability distribution $P$ to another probability distribution $Q$  
\begin{equation}\label{KL}
	D_{KL}(P\|Q) = -\E{P}{ \log\frac{q(x)}{p(x)}} = -\E{P}{ \log{q(x)} } + \E{P}{ \log{p(x)}}, 
\end{equation} 
where $P$ is the generating, or ``true'' distribution of the data, and $Q$ is some approximating distribution. The KL divergence measures the expected difference in negative-log likelihood between distribution $P$ and $Q$, assuming the data was generated by $P$. Due to the similarity of the KL divergence to a negative log-likelihood, we propose the use of the KL divergence as a surrogate for the negative log-likelihood in the GIC (\ref{GIC}). This yields a GIC-type criterion for selecting candidate models generated by the DSS procedure. 

%due to it's similarity to the negative log-likelihood function, which allows us to make use of a number of standard statistical tools when extending the DSS algorithm.  In this paper, we use the Kullback--Leibler (KL) divergence  

In order to use the KL divergence within the GIC framework, we require a reference model and an approximating model. Inspired by the original DSS procedure, we choose the reference model to be the Gaussian distribution with mean $\bar{\by}$ of (\ref{eq:ybar}) and variance $\bar{\sigma}^2 = \E{}{\sigma^2 \vbar \by}$, i.e., the posterior mean of $\sigma^2$. Let $\bbeta_\kappa$ denote the sparse coefficient vector found by solving (\ref{betaDSSg}) or an appropriate approximation to this optimisation problem such as (\ref{DSSnng}); the generalised information score becomes
%%
%\begin{equation}\label{GIC_KLnormal}
%	\mbox{GIC}(\bbeta_\kappa) = \min \limits_{\sigma^2} \left\{ -\E{\tilde\by}{\log{p(\tilde \by \vbar \bbeta_\kappa\sigma^2)}} + \alpha(k_\lambda,\sigma^2) \right\},
%\end{equation} 
%%
%
\begin{equation}\label{GIC_KL}
	\mbox{GIC}(\bbeta_\kappa) = \min \limits_{\sigma^2} \left\{ D(\bar{\bbeta}, \bar{\sigma}^2 \, \| \, \bbeta_{\kappa}, \sigma^2) + \alpha(k_\kappa,\sigma^2) \right\},
\end{equation} 
where 
\begin{equation*}
		D(\bar{\bbeta}, \bar{\sigma}^2 \,  \| \, \bbeta_{\kappa}, \sigma^2) = \frac{n}{2}\log\frac{\sigma^2}{\bar\sigma^2} - \frac{n}{2} + \frac{n^2 \bar\sigma^2}{2\sigma^2} +\frac{\|\bX\bbeta_\kappa-\bX\bar\bbeta  \|_2^2}{2\sigma^2},
\end{equation*} 
is the KL divergence between the reference model, $N(\bX\bar\bbeta,\bar\sigma^2)$ and the candidate sparse model $N(\bX\bbeta_\kappa, \sigma^2)$, and $k_\kappa$ is the degrees of freedom of the sparse regression coefficient vector $\bbeta_\kappa$, which is determined by the particular sparsification scheme that has been employed. Using this approach, the best sparsified regression coefficient vector $\bbeta_\kappa$ is the one that minimises (\ref{GIC_KL}) among all candidate values of $\kappa$. The optimal solution to the group DSS procedure is achieved by balancing the trade-off between the number of predictors in the model and the closeness of the sparse model to the estimated model based on the posterior mean. 

In our definition of the modified GIC (\ref{GIC_KL}) we treat the noise variance parameter $\sigma^2$ of our sparse model as a nuisance parameter and estimate it by minimisation. For a given sparse coefficient vector $\bbeta_\kappa$, the value of $\sigma^2$ that minimises the GIC is given by
%
%In order to compute (), we require an estimate of the noise variance $\sigma^2$ for each candidate sparse model $\bbeta_{\kappa}$. To this end, we treat $\sigma^2$ as a nuisance parameter and set it to the value that minimises the GIC conditional on $\bbeta_{\kappa}$.
%
\begin{equation*}
	\hat \sigma^2 =\frac{ \|\bX\bbeta_\kappa-\bX\bar\bbeta \|^2}{ n - d } + \bar\sigma^2, 
\end{equation*} 
where $d = k_\kappa$ for the MML$_u$ criterion (\ref{MMLu}) and $d=0$ otherwise.

%Although Bayesian samples from posterior samples never produce sparse solution, the plug-in densities have good predictive performance and are easy to compute. They can be evaluated at various points, such as the posterior mean, the posterior median and the posterior mode. In this paper, we choose to use the plug-in density evaluated at the posterior mean as the regular DSS does.

%The KL divergence from normal data to the plug-in density evaluated at the mean of posterior samples in our case becomes:
%\begin{equation}\label{KL_normal}
%D_{\mbox{KL}}=\frac{n}{2}\log\frac{\sigma^2}{\bar\sigma^2} - \frac{n}{2} + \frac{n^2 \bar\sigma^2}{2\sigma^2} +\frac{\|\bX\bbeta_\kappa-\bX\bar\bbeta  \|_2^2}{2\sigma^2}.
%\end{equation} Then it is substituted into (\ref{GIC_KLnormal}) and minimised over $\sigma^2$ to obtain the estimate of $\sigma^2$. For example, the estimates of $\sigma^2$ of information criteria that do not contain $\sigma^2$ in the penalty function $\alpha(\cdot)$ can be derived directly from (\ref{KL_normal}) by differentiating it with respect to $\sigma^2$ and setting to zero to obtain the following:
%\begin{equation*}
%\hat \sigma^2 =\frac{ \|\bX\bbeta_\kappa-\bX\bar\bbeta \|^2}{n}+\bar\sigma^2. 
%\end{equation*} 
%This estimate works for BIC, AIC, AICc in this paper.

This information criteria-based approach has the advantage of allowing us to utilise the substantial body of work that has been undertaken on the problem of variable selection. Further, it provides an objective method for choosing an appropriate degree of posterior sparsification that, in comparison to the heuristic proposed in the original DSS paper, does not require any user-specified parameters. This approach can be used in both the original DSS procedure, as well as in our new group DSS extension to select an appropriate sparse model.

%The group DSS method with the GIC criteria selects a model with a sparse estimate that closely approximates the predictive distribution. 
%Because the mean of posterior samples from the Bayesian methods does not produce sparse estimates, the proposed group DSS method not only selects a sparse estimate, $\bbeta_\kappa$, from the non-negative garrote solution path, but also selects the one that is closet to $\bar\bbeta$ by automatically incorporating the information criteria in the $\kappa$ selection procedure.

%----------------------------------- Subsubsection ----------------------------------- %

\subsubsection{An alternative estimate for the degrees of freedom}

The information criteria approach for selecting a sparse model requires an estimate of the degrees of freedom for all candidate models. In the standard DSS which is based on $\ell_1$ penalisation, the degrees of freedom of each model can be approximated by the number of non-zero regression coefficients. Recall that, in the group DSS procedure proposed in Section~\ref{sec:GDSS}, we approximate the ideal $\ell_0$ minimisation problem (\ref{betaDSSg}) by the more tractable group nonnegative garrotte estimator (\ref{DSSnng}). An estimate for the degrees of freedom (df) of the group NNG was proposed by \cite{2006Yuanp4967}
\begin{equation}
	\label{dfNNG}
	\widehat{\mbox{df}}_{\rm YL} (\bbeta_\kappa) = 2\sum_g\mathbbm{1}(d_g>0) +\sum_g d_g (s_g-2).
\end{equation}
This estimate of the degrees of freedom does not take into account within-group sparsity and will tend to overestimate the effective degrees of freedom for groups where some predictors are inactive. The group DSS procedure uses a smoothed version of the response variable $\by$, $\bar\by$, as the target for the subsequent group NNG sparsification step. Using $\bar{\bbeta}$ in (\ref{DSSnng}) takes into account the coefficient shrinkage induced by the continuous shrinkage prior distributions. Coefficients that are heavily shrunk contribute less than a full degree of freedom; therefore, the effective degrees of freedom for the model are less than suggested by (\ref{dfNNG}) and should be adjusted in accordance with the degree of shrinkage. An overestimation of the degrees of freedom will result in an increased probability of erroneously rejecting large groups that contain only a few active coefficients.

As an illustration, consider a sparse group containing $s_g = 100$ predictors of which only a few are associated with the target. The estimate of the degrees of freedom given by (\ref{dfNNG}) assumes that all $100$ variables within the group are active. However, the target $\bar{\by}$ used in the information criteria is formed from the posterior mean $\bar{\bbeta}$. After shrinkage, unassociated variables within the group will have posterior mean coefficient estimates that are close to zero and have little or no influence on the smoothed data $\bar{\by}$. This means that the information criteria penalty terms for models containing large groups where some predictors are inactive will be much larger than if the posterior shrinkage was taken into account.

 %which includes variables that may have little to no contribution to the target. nd 
%therefore produces a higher value of the degrees of freedom than is warranted. This leads to a larger value in information criteria and less chance of selecting the model with this group. However, this group is active and should be selected. 

We propose an alternative estimate of the degrees of freedom that takes into account the degree of coefficient shrinkage within all groups. The key idea is to exploit the form of the Bayesian linear regression hierarchy (\ref{BGLSRM}) which allows us to calculate the posterior expected degrees of freedom. Recall that the hierarchical representation of the global-local shrinkage model is 
\begin{align*}
\by \, | \, \bX,\bbeta,\sigma^2 & \, \sim \,  N(\bX\bbeta,\sigma^2\mathbf{I}_n),  \\
\bbeta \, | \, \sigma^2,\tau^2,\blambda,\bdelta & \, \sim  \, N(\boldsymbol{0},\sigma^2\tau^2\bD_{\blambda}\bD_{\bdelta_1}\cdots\bD_{\bdelta_K}).
\end{align*}
Conditional on $\tau$, $\bdelta$ and $\blambda$, the above hierarchy reduces to a generalised ridge regression model. The posterior mean of $\bbeta$ for this hierarchy is
\begin{equation}\label{ourridge}
\bar{\bbeta}_{\tau^2, \blambda, \bdelta} = \Exx \left[ \bbeta \vbar \by, \tau^2, \blambda, \bdelta \right] = \left(\bX^T\bX + \bSigma_{\tau^2, \blambda, \bdelta} \right)^{-1} \bX^T \by
\end{equation}
where $\bSigma_{\tau^2, \blambda, \bdelta} = (\tau^2 \bD_{\blambda} \bD_{\bdelta_1} \cdots \bD_{\bdelta_K})^{-1}$ and the matrices $\bD_{\blambda}$ and $\bD_{\bdelta_k}$ are defined in Section~\ref{sec:BGGLSHM}.
The effective degrees of freedom of a ridge regression model, conditional on the hyper-parameters $\tau^2, \blambda, \bdelta$, is~\citep{2001Friedmanp}
\begin{equation}
	\label{eq:dfRidge}
	\mbox{df}(\bar{\bbeta}_{\tau^2, \blambda, \bdelta}) =  \mbox{tr}\left( \bX (\bX^T\bX+\bSigma_{\tau^2, \blambda, \bdelta})^{-1}\bX^T \right).
\end{equation} 
This estimate of the degrees of freedom depends on the particular values of the hyper-parameters $\tau^2$, $\blambda$ and $\bdelta$; to remove this dependency we average (\ref{eq:dfRidge}) over the posterior distribution of $\tau^2$, $\blambda$ and $\bdelta$ which yields the posterior expected degrees of freedom:
\begin{equation}
	\label{eq:e:dfRidge}
	\mbox{df}(\bar{\bbeta}) =  {\mathbb E}_{\tau^2, \blambda, \bdelta} \left[ \mbox{tr}\left( \bX (\bX^T\bX+\bSigma_{\tau^2, \blambda, \bdelta})^{-1}\bX^T \right) \vbar \by \right].
\end{equation} 
This equation yields an estimate for the degrees of freedom of the full regression model which can be used to compute the information criterion score for any candidate model in the sparsification path produced by the group NNG. However, when the number of candidate models is large, this approach is computationally challenging. An approximation to (\ref{eq:e:dfRidge}) that is less computationally expensive can be found by decomposing the degrees of freedom of the full regression model into the sum of the degrees of freedom of each of the $G$ groups. In the case that all groups are mutually orthogonal, this approximation is equivalent to the degrees of freedom of the full regression model (\ref{eq:e:dfRidge}). We therefore require an estimate of the degrees of freedom for each of the $G$ groups that comprise the full model. As discussed in Section~\ref{sec:GDSS}, we restrict our attention to models with a single level of $G$ non-overlapping groups. In this case, the degrees of freedom for a group $g$ is
%However, in order to sparsify the posterior in terms of groups we require an estimate of the degrees of freedom for each of the individual groups. This may be obtained  
%
\begin{equation}
	\label{eq:dfestg}
	\mbox{df}(\bar{\bbeta}_g)  = \Exx\left[\mbox{tr}\left(\bX_g \left(\bX_g^T\bX_g+ \bSigma_g \right)^{-1}\bX_g^T \right) \vbar \by \right],
\end{equation} 
where $\bSigma_g =  (\tau^2{\bD_{\blambda}}_g {\bD_{\bdelta}}_g )^{-1}$. Using (\ref{eq:dfestg}), our estimate for the degrees of freedom for a candidate model $\bbeta_{\kappa}$ in the group NNG path is 
\begin{equation}\label{dfest}
\widehat{\mbox{df}}_{\rm PE}(\bbeta_{\kappa}) = \sum\limits_{g=1}^G \mathbbm{1}(d_g\neq0)\cdot \mbox{df}(\bar{\bbeta}_g) . % \Exx\left[\mbox{tr}\left(\bX_g'  \bX_g \left(\bX_g'\bX_g+ \bSigma_g \right)^{-1}\right) \right].
\end{equation}
In practice, to calculate the expectation in (\ref{eq:dfestg}), we use the samples from the posterior distribution and evaluate the equivalent expression
\begin{equation}
	\label{dfestg}
	\mbox{df}(\bar{\bbeta}_g)  = \mbox{tr}\left(\bX_g^T \bX_g \,  \Exx \left[  \left(\bX_g^T\bX_g+ \bSigma_g \right)^{-1} \vbar \by \right] \right) ,	
\end{equation} 
which requires only one evaluation of $\bX_g^T \bX_g$, and involves less computation than~\eqref{eq:dfestg}.

%To compute The hyper-parameters are unknown and must be estimated from the data. We in
%We To remove the dependency of $\mbox{df}(\bar{\bbeta}_{\tau^2, \blambda, \bdelta})$ on the unknown values of the hyper-parameters, we  
%
%Comparing (\ref{ourridge}) and (\ref{ridge}), an estimate of df for group $g,g\in\{1,\cdots,G\}$ can be derived as:
%
%\begin{equation}
	%\label{dfestg}
	%\mbox{df}(\bar{\bbeta}_g)  = \Exx\left[\mbox{tr}\left(\bX_g \left(\bX_g'\bX_g+ \bSigma_g \right)^{-1}\bX_g' \right) \vbar \by \right],
%\end{equation} 
%%
%where $\bSigma_g = (\tau^2{\bD_{\blambda}}_g {\bD_{\bdelta_1}}_g \cdots {\bD_{\bdelta_K}}_g )^{-1}$. Therefore, the approximation to the df for the group non-negative garrotte can be proposed as:
%%
%\begin{equation}\label{dfest}
%\widehat{\mbox{df}}= \sum\limits_{g=1}^G \mathbbm{1}(d_g\neq0)\cdot \Exx\left[\mbox{tr}\left(\bX_g'  \bX_g \left(\bX_g'\bX_g+ \bSigma_g \right)^{-1}\right) \right].
%\end{equation}
%
%All posterior samples have been obtained in the sampling procedure, so it is straightforward to calculate $\bSigma_g$ and the estimate of df. However, if $\bX_g$ are not mutually orthogonal, the expectation of the trace tends to grow and the estimate of df increases as well. 
%
%In the special case where there is only one level of group structure, the $\bSigma_g$ in the above proposed df equation becomes:
%\begin{equation}\label{dfest1}
%\bSigma_g=  (\tau^2{\bD_{\blambda}}_g {\bD_{\bdelta}}_g )^{-1}.
%\end{equation}

%================================================%
%===================== Section =====================%
\section{Results and Discussion}

%----------------------------------- Subsection ----------------------------------- %
\subsection{Example 1}
We compared the performance of our Bayesian sparse group selection using continuous shrinkage priors against the Bayesian group spike-and-slab approach (BSGS) in \cite{2016Chenp665683}. The BSGS approach has been shown to outperform the standard group variable selection method, the sparse group Lasso~\citep{2010Friedmanp}, in simulation studies, and can be viewed as a potential ``gold standard'' in Bayesian grouped inference. Our simulations followed the experimental setup of \cite{2016Chenp665683}. In Example~1 there were $n=50$ observations and $p=60$ predictors. The predictors were divided into six groups, in which the first two groups contained 5 variables each, the second two groups contained 10 variables each and the last two groups contained 15 variables each. There were a total of 5 active predictors with coefficients $\beta_3=3.2$, $\beta_{11}=-2$, $\beta_{12}=1$, $\beta_{31}=1.5$, $\beta_{32}=-1.5$. The design matrix, $\bX$, was generated from the multivariate normal distribution with a mean of 0, a variance of 1.25, and a correlation of 0.2 between any two variables in the same group and 0 otherwise. The variance of the errors, $\sigma^2$, was varied between 4 to 2, resulting in a signal-to-noise ratio of 6.74 and 12.49, respectively.

We used the grouped Bayesian horseshoe prior (see Section~\ref{sec:BGHSM}) to obtain 10,000 posterior samples with the first 1,000 samples discarded as burn-in. The sampling procedure was done using the `BayesReg' software toolbox~\citep{2016EnespHDBRRWTBP} which implements the grouped global-local shrinkage priors detailed in this paper; this can be downloaded from MathWorks File Exchange (ID: 60823). The mean of the posterior samples was then used in the group non-negative garrote to obtain the solution path $\bbeta_\kappa$. Four information criteria, BIC, AIC, AIC$_c$ and MML$_u$ (see Section~\ref{sec:ICGDSS}), were computed for each model in the path using two different estimates of the degrees of freedom: (1)~the conventional group NNG degrees of freedom given by~(\ref{dfNNG}), and (2)~the posterior expected degrees of freedom given by~(\ref{dfest}), which we called BIC$^*$, AIC$^*$, AIC$_c^*$ and MML$_u^*$. For the AIC$_c$ criterion, if the estimated degrees of freedom is greater than the number of observations $n$, the denominator of the second term in the AICc formula (\ref{eq:aicc}) is negative and the corresponding candidate models are not considered for selection. The final sparse estimates were chosen by minimisation of the four information criteria. As a comparison, we also calculated the variation explained and excess error for each sparse model in the grouped NNG path (see Section~\ref{sec:DSSM}). We selected the sparsest model for which the 90\% credible interval of the variation explained and excess error statistics contained the expected value of the variation explained and excess error when $\kappa$ is zero~\citep{2015Hannp435448}, respectively.

Since the datasets used for this example in \cite{2016Chenp665683} were not provided, we instead generated $10,000$ datasets using the procedure described above and recorded the number of times each method correctly identified groups as active or inactive. Due to the very long run times of the BSGS procedure, we did not include the BSGS method in these simulations; instead, we used the results reported in \cite{2016Chenp665683} for comparison and scale our results to match the number of simulations in \cite{2016Chenp665683}.

%and in order to make our results comparable to the BSGS method, we divided the our results by 100 so that it is on the same scale of the BSGS.

The results for the simulations in Example~1, when $\sigma^2=4$, are shown in Table~\ref{ex1_sigma24}. In terms of the overall sum of correct identifications, the methods with estimated degrees of freedom $\widehat{\mbox{df}}_{\rm YL}$ frequently failed to select active sparse groups (i.e., most predictors within the group are unassociated with the target) that were associated with the target. In contrast, our proposed degrees of freedom $\widehat{\mbox{df}}_{\rm PE}$ showed a great improvement compared with $\widehat{\mbox{df}}_{\rm YL}$ for BIC, AIC$_c$ and MML$_u$. This was particularly apparent for group 5, which is an extremely sparse group with only two of the 15 predictors active. Our proposed method produced a much smaller estimate of the degrees of freedom and therefore resulted in higher probabilities of correctly selecting active sparse groups.

Although AIC with $\widehat{\mbox{df}}_{\rm PE}$ (i.e., AIC$^*$ in Table~\ref{ex1_sigma24}) correctly identified more active sparse groups than AIC with $\widehat{\mbox{df}}_{\rm YL}$, it selected more inactive groups; as the AIC is known to overfit, particularly for small sample sizes $n$ or when the number of predictors, $p$, is close to $n$, this increased rate of false positives is likely due to the AIC approximations being inaccurate in this setting. The AIC$_c$ using $\widehat{\mbox{df}}_{\rm YL}$ did not perform well at selecting the active sparse groups in this example as it tended to overestimate the degrees of freedom associated with these groups. However, AIC$_c$ with our proposed estimate $\widehat{\mbox{df}}_{\rm PE}$ appeared to more accurately estimate the effective degrees of freedom, and therefore tended to correctly select the active sparse groups more frequently. The BIC$^*$, AICc$^*$ and MMLu$^*$ had very competitive performance in comparison with the original BSGS procedure, while being substantially quicker to run.

%When $\sigma^2 = 2$, i.e. a larger signal-to-noise ratio, all methods except AICc using $\widehat{\mbox{df}}_{\rm YL}$ performed similarly (see Table~\ref{ex1_sigma22}). In this case, the noise is relatively small, which makes true active groups easy to identify. Using the posterior expected degrees of freedom $\widehat{\mbox{df}}_{\rm PE}$ showed a slight improvement in terms of overall sum of correct identifications compared to using the standard estimate $\widehat{\mbox{df}}_{\rm YL}$.

%-------------------------- Ex 1 sigma2=4 -----------------------------%
\begin{table}[h]
%\scriptsize
\centering
\caption{Percentage of times the six groups are selected in Example 1 for $\sigma^2 = 4$ by Bayesian sparse group selection model (BSGS), variation explained (VarExp), excess error (ExcErr), information criteria approaches with degrees of freedom $\widehat{\mbox{df}}_{\rm YL}$ (BIC, AIC, AIC$_c$, MML$_u$), and information criteria approaches with posterior expected degrees of freedom $\widehat{\mbox{df}}_{\rm PE}$ (BIC$^*$, AIC$^*$, AIC$_c^*$, MML$_u^*$). Active groups are marked in bold.}
\label{ex1_sigma24}
\resizebox{\textwidth}{!} {%
\begin{tabular}{ccccccccccccccc}
\toprule
Group & \begin{tabular}[c]{@{}c@{}} Active No. of Variables\\/Group Size\end{tabular}        & BSGS & VarExp & ExcErr &~& BIC & AIC & AIC$_c$ & MML$_u$&~&BIC$^*$&AIC$^*$&AIC$_c^*$&MML$_u^*$ \\ 
\cmidrule{1-15}

\textbf{1} & 1/5 & 100 & 100.0 & 100.0 & & 99.7 & 100.0 & 100.0 & 100.0 & & 100.0 & 100.0 & 100.0 & 100.0  \\ 
2 & 0/5 & 7 & 1.8 & 2.5 & & 0.4 & 2.8 & 0.1 & 0.2 & & 8.4 & 18.6 & 8.5 & 6.1  \\ 
\textbf{3} & 2/10 & 100 & 97.6 & 97.3 & & 66.5 & 96.9 & 65.8 & 77.1 & & 99.3 & 99.8 & 99.6 & 99.1  \\ 
4 & 0/10 & 1 & 1.4 & 2.3 & & 0.1 & 1.7 & 0.0 & 0.0 & & 5.5 & 14.8 & 5.2 & 3.5  \\ 
\textbf{5} & 2/15 & 97 & 89.7 & 90.3 & & 39.9 & 87.3 & 7.1 & 41.7 & & 95.8 & 98.5 & 97.3 & 94.8  \\ 
6 & 0/15 & 2 & 1.1 & 2.0 & & 0.1 & 0.9 & 0.0 & 0.0 & & 3.9 & 12.0 & 3.4 & 2.1  \\ 

\cmidrule{1-15}
\multicolumn{2}{c}{\begin{tabular}[c]{@{}c@{}}Overall sum of\\   correct identifications\end{tabular}}  &  587 & 583.0 & 580.9 & & 505.6 & 578.8 & 472.8 & 518.6 & & 577.2 & 552.8 & 579.8 & 582.1  \\ 

\bottomrule
\end{tabular}%
}
\end{table}
%-------------------------- End of Ex 1 sigma2=4-----------------------------%

The results for $\sigma^2 = 2$, i.e. a larger signal-to-noise ratio, are shown in Table~\ref{ex1_sigma22}. In this case, the level of noise is low and the active groups are easier to correctly identify. From the table, the MML and AIC$_c$ criteria using $\widehat{\mbox{df}}_{\rm PE}$ showed better performance in selecting sparse groups (group 3 and 5) compared to the same criteria using $\widehat{\mbox{df}}_{\rm YL}$. In terms of the overall sum of correct identifications, the MML$_u^*$, variation explained and excess error criteria were all virtually indistinguishable from the BSGS method, with AIC$_c^*$ being only slightly worse. The BIC and AIC criteria with $\widehat{\mbox{df}}_{\rm YL}$ performed better than the same criteria with $\widehat{\mbox{df}}_{\rm PE}$; this appears to be due the inaccuracy of the asymptotic approximations used in these methods, meaning that the conservative estimate $\widehat{\mbox{df}}_{\rm YL}$ of the degrees of freedom artificially assisted in preventing overfitting. This was supported by the fact that the small sample size versions of these criteria (i.e., MML$_u^*$ and AIC$_c^*$) performed substantially better with our proposed estimate of the degrees of freedom.
%

%----------------------------- Ex1 sigma2=2 -----------------------------%

\begin{table}[h]
%\scriptsize
\centering
\caption{Percentage of times the six groups are selected in Example 1 for $\sigma^2 = 2$ by Bayesian sparse group selection model (BSGS), variation explained (VarExp), excess error (ExcErr), information criteria approaches with degrees of freedom $\widehat{\mbox{df}}_{\rm YL}$ (BIC, AIC, AIC$_c$, MML$_u$), and information criteria approaches with posterior expected degrees of freedom $\widehat{\mbox{df}}_{\rm PE}$ (BIC$^*$, AIC$^*$, AIC$_c^*$, MML$_u^*$). Active groups are marked in bold.}
\label{ex1_sigma22}
\resizebox{\textwidth}{!} {%
\begin{tabular}{ccccccccccccccc}
\toprule
Group&\begin{tabular}[c]{@{}c@{}} Active No. of Variables\\/Group Size\end{tabular}        & BSGS & VarExp & ExcErr &~& BIC & AIC & AICc & MML$_u$&~&BIC$^*$&AIC$^*$&AIC$_c^*$&MML$_u^*$ \\ 
\cmidrule{1-15}

\textbf{1} & 1/5 & 100 & 100.0 & 100.0 & & 100.0 & 100.0 & 100.0 & 100.0 & & 100.0 & 100.0 & 100.0 & 100.0  \\ 
2 & 0/5 & 2 & 1.0 & 1.8 & & 0.7 & 2.6 & 0.0 & 0.0 & & 6.8 & 16.8 & 6.0 & 3.2  \\ 
\textbf{3} & 2/10 & 100 & 100.0 & 100.0 & & 98.0 & 100.0 & 79.6 & 95.6 & & 100.0 & 100.0 & 100.0 & 100.0  \\ 
4 & 0/10 & 1 & 0.8 & 1.7 & & 0.1 & 1.7 & 0.0 & 0.0 & & 4.6 & 13.7 & 3.7 & 1.8  \\ 
\textbf{5} & 2/15 & 100 & 99.7 & 99.8 & & 94.1 & 99.7 & 20.6 & 82.0 & & 99.9 & 100.0 & 100.0 & 99.8  \\ 
6 & 0/15 & 0 & 0.7 & 1.6 & & 0.1 & 0.9 & 0.0 & 0.0 & & 3.3 & 11.5 & 2.0 & 0.8  \\ 

\cmidrule{1-15}
\multicolumn{2}{c}{\begin{tabular}[c]{@{}c@{}}Overall sum of\\   correct identifications\end{tabular}}   & 597  & 597.1 & 594.6 & & 591.2 & 594.4 & 500.2 & 577.5 & & 585.3 & 557.9 & 588.2 & 594.0  \\ 
\bottomrule
\end{tabular}%
}
\end{table}
%----------------------------- End of Ex1 sigma2=2 -----------------------------%

For this example, we ran further simulations for $\sigma^2 =8$ and $\sigma^2 =16$ (results not shown). When the level of noise was high, our proposed methods with $\widehat{\mbox{df}}_{\rm PE}$ generally selected active sparse groups more frequently than the same criteria using $\widehat{\mbox{df}}_{\rm YL}$, and the AIC$_c^*$ and MML$_u^*$ criteria produced a greater overall sum of correct identifications compared to the other methods tested, including the variation explained and excess error criteria. From all the criteria tested, the MML$_u^*$ criterion appeared to have the best overall performance, in terms of speed and accuracy.

%----------------------------------- Subsection ----------------------------------- %
\subsection{Example 2}
In Example~2, there were $n=200$ observations and $p=100$ predictors divided into 10 groups of equal size. The number of non-zero coefficients in the first six groups was 10, 8, 6, 4, 2, and 1, respectively. The non-zero coefficients were randomly sampled from the set $\{ -1, 1 \}$. The noise was sampled from independent and identically distributed normal distributions with mean zero and variance $16$. In this case, the signal-to-noise ratio was 3.42. We ran the example for 100 iterations using the data provided by \cite{2016Chenp665683} and recorded the number of times that each group was correctly identified. The results are presented in Table~\ref{ex2}.

%Comparing the results from Table~\ref{ex2}, our method with the proposed estimates of df showed a substantial improvement in correct identifications over the df from NNG for BIC, AICc and MMLu. In particular, the correct identifications for sparse groups (G5 and G6) using our proposed estimate of df are much more than using the estimate of df from NNG. 

%----------------------------- Ex2 -----------------------------%
\begin{table}[h]
%\scriptsize
\centering
\caption{Number of times the ten groups are selected in Example 2 by Bayesian sparse group selection model (BSGS), variation explained (VarExp), excess error (ExcErr), information criteria approaches with degrees of freedom $\widehat{\mbox{df}}_{\rm YL}$ (BIC, AIC, AIC$_c$, MML$_u$), and information criteria approaches with posterior expected degrees of freedom $\widehat{\mbox{df}}_{\rm PE}$ (BIC$^*$, AIC$^*$, AIC$_c^*$, MML$_u^*$). Active groups are marked in bold.}
\label{ex2}
\resizebox{\textwidth}{!} {%
\begin{tabular}{ccccccccccccccc}
\toprule
Group&\begin{tabular}[c]{@{}c@{}} Active No. of Variables\\/Group Size\end{tabular}        & BSGS & VarExp & ExcErr &~& BIC & AIC & AICc & MML$_u$&~&BIC$^*$&AIC$^*$&AIC$_c^*$&MML$_u^*$ \\ 
\cmidrule{1-15}

%\textbf{1} & 10/10 & 100 & 100 & 100 & & 100 & 100 & 100 & 100 & & 100 & 100 & 100 & 100  \\ 
%\textbf{2} & 8/10 & 100 & 100 & 100 & & 96 & 100 & 100 & 100 & & 100 & 100 & 100 & 100  \\ 
%\textbf{3} & 6/10 & 100 & 96 & 99 & & 88 & 100 & 98 & 96 & & 100 & 100 & 100 & 100  \\ 
%\textbf{4} & 4/10 & 99 & 90 & 96 & & 72 & 96 & 93 & 93 & & 98 & 100 & 99 & 99  \\ 
%\textbf{5} & 2/10 & 94 & 53 & 70 & & 26 & 71 & 63 & 55 & & 80 & 88 & 87 & 84  \\ 
%\textbf{6} & 1/10 & 79 & 20 & 48 & & 3 & 44 & 24 & 22 & & 58 & 73 & 66 & 61  \\ 
%7 & 0/10 & 10 & 1 & 3 & & 0 & 3 & 1 & 1 & & 4 & 15 & 7 & 5  \\ 
%8 & 0/10 & 12 & 2 & 4 & & 0 & 4 & 2 & 2 & & 5 & 11 & 8 & 6  \\ 
%9 & 0/10 & 11 & 0 & 2 & & 0 & 1 & 0 & 0 & & 3 & 18 & 10 & 8  \\ 
%10 & 0/10 & 14 & 0 & 0 & & 0 & 0 & 0 & 0 & & 4 & 17 & 6 & 4  \\ 
\textbf{1} & 10/10 & 100 & 100 & 100 & & 100 & 100 & 100 & 100 & & 100 & 100 & 100 & 100  \\ 
\textbf{2} & 8/10 & 100 & 100 & 100 & & 96 & 100 & 100 & 100 & & 100 & 100 & 100 & 100  \\ 
\textbf{3} & 6/10 & 100 & 96 & 99 & & 88 & 99 & 97 & 96 & & 99 & 100 & 100 & 100  \\ 
\textbf{4} & 4/10 & 99 & 90 & 96 & & 72 & 96 & 93 & 93 & & 97 & 100 & 99 & 99  \\ 
\textbf{5} & 2/10 & 94 & 53 & 70 & & 26 & 71 & 64 & 56 & & 79 & 88 & 87 & 83  \\ 
\textbf{6} & 1/10 & 79 & 19 & 46 & & 3 & 45 & 24 & 21 & & 57 & 73 & 66 & 62  \\ 
7 & 0/10 & 10 & 1 & 3 & & 0 & 3 & 1 & 1 & & 3 & 14 & 7 & 5  \\ 
8 & 0/10 & 12 & 2 & 4 & & 0 & 4 & 2 & 2 & & 6 & 12 & 8 & 6  \\ 
9 & 0/10 & 11 & 0 & 2 & & 0 & 1 & 0 & 0 & & 3 & 18 & 10 & 7  \\ 
10 & 0/10 & 14 & 0 & 0 & & 0 & 0 & 0 & 0 & & 4 & 15 & 6 & 4  \\ 

\cmidrule{1-15}
\multicolumn{2}{c}{\begin{tabular}[c]{@{}c@{}}Overall sum of\\   correct identifications\end{tabular}} & 925  &  855  & 902 &&  785 &  903  & 875  & 863 &&  916 &  902  & 921 &  922\\
\bottomrule
\end{tabular}%
}
\end{table}
%

%----------------------------- End of Ex2 -----------------------------%
The BIC, AIC$_c$ and MML$_u$ criteria using the posterior expected degrees of freedom $\widehat{\mbox{df}}_{\rm PE}$ show improved rates of selection of the sparse active groups, and are substantially better in terms of overall numbers of correct identifications, in comparison to the same criteria using the conventional degrees of freedom $\widehat{\mbox{df}}_{\rm YL}$. The overall sum of correct identifications for BIC$^*$, AIC$_c^*$ and particularly MML$_u^*$, were similar to the results obtained by the BSGS method. While BIC$^*$, AIC$_c^*$ and MML$_u^*$ produced a competitive overall sum of correct identifications, they also picked less inactive groups than the BSGS did. Both the excess error, and particularly the variance explained criteria appear inferior to BSGS and BIC$^*$, AICc$^*$ and MMLu$^*$ in this setting. These criteria appear overly conservative, and achieve a low rate of correct identification of the sparsest groups (5 and 6).

While the results of the proposed group DSS procedure were highly competitive compared to the slab-and-spike BSGS approach in terms of the correct identifications of the groups, the computational cost of our procedure was significantly lower. For example, it took approximately 30 hours to run the BSGS for a single simulation with the prior variance for active variables $\tau$ varied over $\{0.5,1\}$; in contrast, our group DSS method using the grouped horseshoe prior took approximately 5 seconds to finish a single simulation, and approximately 500 seconds to complete all 100 simulations.

%----------------------------------- Subsection ----------------------------------- %
\subsection{Real data -- birth weight dataset}
We compared our proposed group DSS procedure to the BSGS method using the birth weight dataset from~\cite{1989Homserp}. The data was collected by the Baystate Medical Center, Springfield, Massachusetts, during 1986. There were $n=189$ observations and $p=8$ predictors. The response variable was birth weight measured in grams. The eight predictors included mother's age (AGE), weight in pounds at the last menstrual period (LWT), race (RACE: white, black and others), smoking status during pregancy (SMOKE: yes or no), history of premature labor (PTL: 0, 1, 2, etc.), history of hypertension (HT: yes or no), presence of uterine irritability (UI: yes or no), and number of physician visits during the first trimester (FTV: 0, 1, 2, etc.). We adopted the same procedure as \cite{2006Yuanp4967} and expanded the two continuous variables AGE and LWT using third-order polynomials treating each as a separate group. The categorical variable RACE contained three factors and was turned into two dummy variables, which also formed a group. The rest of the variables were unaltered and formed separate singleton groups. We used our grouped DSS procedure with the grouped horseshoe prior to select several sparse models using the four information criteria discussed in Section~\ref{sec:ICGDSS}. The models selected by all four information criteria using the posterior expected degrees of freedom~(\ref{dfest}) were identical. The final model included five groups
\begin{equation}
\begin{split}
\Exx(Y|X) = &\mbox{LWT} + \mbox{LWT}^2 + \mbox{LWT}^3 + \mbox{RACE}_{\mbox{black-white}} + \mbox{RACE}_{\mbox{others-white}}\\
& + \mbox{SMOKE} + \mbox{HT} + \mbox{UI}.
\end{split}
\end{equation}

This same dataset was previously analysed by~\cite{2010Farcomenip10431062} whose final model included LWT, LWT$^2$, RACE, SMOKE, HT, UI, and HT$:$RACE, where HT$:$RACE is an interaction term. In the original paper of \cite{2016Chenp665683}, interaction terms were not included in the analysis; to make our experiments comparable to the experiment in the \cite{2016Chenp665683}, we omitted all interaction terms in our analysis. The BSGS method was also used to investigate the same dataset and selected six groups: LWT, LWT$^2$, LWT$^3$, $\mbox{RACE}_{\mbox{black-white}}$, $\mbox{RACE}_{\mbox{others-white}}$, SMOKE, HT, UI, and PTL, with the posterior probability of PTL (0.512) barely passing the threshold value 0.5.

We also compared our method with the BSGS procedure in terms of predictive performance. We divided the dataset into two parts, a training dataset (75\% of samples) and a testing dataset (25\% of samples), and ran 100 simulations using the two algorithms. The training data were used to find a sparse estimate and the testing data were then used to compute square prediction errors for the estimates obtained from the training data. For the BSGS method, the best $\tau$ was selected from the set $\{1.5,2,2.5,3\}$ for each of the training datasets.

The ratios of the mean-squared prediction errors for each method, all relative to the mean-squared prediction error achieved by the BSGS procedure, are shown in Table~\ref{petable2}. The posterior mean estimates $\bar\bbeta$ produced the best prediction error of all the methods tested; however, the posterior mean estimate is never sparse. The prediction errors obtained using the variance explained and excess error criteria are essentially identical to those obtained by BSGS. In contrast, our proposed criteria performed better than the BSGS, variance explained and excess error procedures. %Interestingly, in addition to producing sparse estimates, the AIC$^*$, AIC$_c^*$ and MML$_u^*$ criteria performed almost as well as the posterior mean in terms of prediction error.

%The ratios of our four methods were all smaller than one, so they all produced smaller mean-squared prediction errors compared with the BSGS. Therefore, our proposed methods had a better predictive performance.

%----------------------------- Table Real Data -----------------------------%
\begin{table}[h]
\centering
\caption{The ratio of mean-squared prediction errors for each method relative to the BSGS procedure for the birth weight data.}
\label{petable2}
\begin{tabular}{cccccccc}
\toprule
Methods   & VarExp   & ExcErr   & BIC$^*$      & AIC$^*$      & AIC$_c^*$     & MML$_u^*$     & $\bar\bbeta$ \\ 
\cmidrule{2-8}
Ratio	&	1.002	&0.984	&0.979 &	0.942&	0.946&	0.940 &0.906\\
\bottomrule
\end{tabular}
\end{table}
%----------------------------- End of Table Real Data -----------------------------%

%===================== Section =====================%
\section{Conclusion}
In this paper, we proposed a group Bayesian global-local shrinkage hierarchical model which incorporates overlapping and multilevel group structures. We extended two shrinkage priors, the horseshoe and the horseshoe+ priors, to the setting of grouped variables. We also adapted the decoupled shrinkage and selection (DSS) method to handle grouped regression models, and use the grouped non-negative garrotte to sparsify posterior estimates. To select a final sparse model, we combined four information criteria approaches with the grouped DSS procedure and proposed an improved estimate of the degrees of freedom for a candidate model in the grouped non-negative garotte solution path. As discussed in Section~\ref{sec:GDSS}, our group DSS procedure can be extended to select groups for multilevels, however, the detail is beyond the scope of this paper and will be considered in future work. 

Simulation results showed that the procedure using our proposed estimate of the degrees of freedom performed well in selecting active groups, especially when these groups were sparse. Our proposed method demonstrated similar performance to the Bayesian grouped slab-and-spike (BSGS) method in terms of the overall rate of correct group identifications, while requiring substantially less computational resources. In all simulations, the use of small-sample information criteria to select sparse models appeared to be preferable to the original heuristic model selection criteria proposed in~\cite{2015Hannp435448}. The prediction errors of our proposed method obtained in an experiment using real data also showed an improvement over the BSGS method and the original DSS model selection criteria. These results suggest that the proposed grouped DSS procedure with information criteria is an efficient tool for selecting sparse grouped models with good predictive performance.

%===================== Section =====================%

\label{Bibliography}
\bibliographystyle{Chicago}
%\bibliography{papers_in_jabref}

\begin{thebibliography}{}

\bibitem[\protect\citeauthoryear{Akaike}{Akaike}{1973}]{1973Akaikep267281}
Akaike, H. (1973).
\newblock Information theory and an extension of the maximum likelihood
  principle.
\newblock {\em Second International Symposium on Information Theory\/}~{\em I},
  267--281.

\bibitem[\protect\citeauthoryear{Bayarri and Berger}{Bayarri and
  Berger}{2004}]{2004Bayarrip5880}
Bayarri, M.~J. and J.~O. Berger (2004).
\newblock The interplay of {B}ayesian and frequentist analysis.
\newblock {\em Statistical Science\/}~{\em 19\/}(1), 58--80.

\bibitem[\protect\citeauthoryear{Bhadra, Datta, Polson, and Willard}{Bhadra
  et~al.}{2016}]{2016Bhadrap}
Bhadra, A., J.~Datta, N.~G. Polson, and B.~Willard (2016).
\newblock The horseshoe+ estimator of ultra-sparse signals.
\newblock {\em Bayesian Analysis\/}.

\bibitem[\protect\citeauthoryear{Bhattacharya, Pati, Pillai, and
  Dunson}{Bhattacharya et~al.}{2015}]{2015Bhattacharyap14791490}
Bhattacharya, A., D.~Pati, N.~S. Pillai, and D.~B. Dunson (2015).
\newblock Dirichlet--{L}aplace priors for optimal shrinkage.
\newblock {\em Journal of the American Statistical Association\/}~{\em
  110\/}(512), 1479--1490.

\bibitem[\protect\citeauthoryear{Bondell and Reich}{Bondell and
  Reich}{2012}]{2012Bondellp16101624}
Bondell, H.~D. and B.~J. Reich (2012).
\newblock Consistent high-dimensional {B}ayesian variable selection via
  penalized credible regions.
\newblock {\em Journal of the American Statistical Association\/}~{\em
  107\/}(500), 1610--1624.

\bibitem[\protect\citeauthoryear{Carvalho, Polson, and Scott}{Carvalho
  et~al.}{2009}]{2009Carvalhop7380}
Carvalho, C.~M., N.~G. Polson, and J.~G. Scott (2009).
\newblock Handling sparsity via the horseshoe.
\newblock In D.~van Dyk and M.~Welling (Eds.), {\em Proceedings of Machine
  Learning Research}, Volume~5, pp.\  73--80.

\bibitem[\protect\citeauthoryear{Carvalho, Polson, and Scott}{Carvalho
  et~al.}{2010}]{2010Carvalhop465480}
Carvalho, C.~M., N.~G. Polson, and J.~G. Scott (2010).
\newblock The horseshoe estimator for sparse signals.
\newblock {\em Biometrika\/}~{\em 97\/}(2), 465--480.

\bibitem[\protect\citeauthoryear{Chen, Chu, Yuan, and Wu}{Chen
  et~al.}{2016}]{2016Chenp665683}
Chen, R.-B., C.-H. Chu, S.~Yuan, and Y.~N. Wu (2016).
\newblock Bayesian sparse group selection.
\newblock {\em Journal of Computational and Graphical Statistics\/}~{\em
  25\/}(3), 665--683.

\bibitem[\protect\citeauthoryear{Farcomeni}{Farcomeni}{2010}]{2010Farcomenip10431062}
Farcomeni, A. (2010).
\newblock Bayesian constrained variable selection.
\newblock {\em Statistica Sinica\/}~{\em 20}, 1043--1062.

\bibitem[\protect\citeauthoryear{Friedman, Hastie, and Tibshirani}{Friedman
  et~al.}{2001}]{2001Friedmanp}
Friedman, J., T.~Hastie, and R.~Tibshirani (2001).
\newblock {\em The elements of statistical learning}.
\newblock Springer series in statistics New York.

\bibitem[\protect\citeauthoryear{Friedman, Hastie, and Tibshirani}{Friedman
  et~al.}{2010}]{2010Friedmanp}
Friedman, J., T.~Hastie, and R.~Tibshirani (2010).
\newblock A note on the group {L}asso and a sparse group {L}asso.
\newblock {\em arXiv preprint arXiv:1001.0736\/}.

\bibitem[\protect\citeauthoryear{George and Mc{C}ulloch}{George and
  Mc{C}ulloch}{1993}]{1993Georgep881889}
George, E.~I. and R.~E. Mc{C}ulloch (1993).
\newblock Variable selection via {G}ibbs sampling.
\newblock {\em Journal of the American Statistical Association\/}~{\em
  88\/}(423), 881--889.

\bibitem[\protect\citeauthoryear{Giurc{\u{a}}neanu, Razavi, and
  Liski}{Giurc{\u{a}}neanu et~al.}{2011}]{2011Giurcaneanup16711692}
Giurc{\u{a}}neanu, C.~D., S.~A. Razavi, and A.~Liski (2011).
\newblock Variable selection in linear regression: Several approaches based on
  normalized maximum likelihood.
\newblock {\em Signal Processing\/}~{\em 91\/}(8), 1671--1692.

\bibitem[\protect\citeauthoryear{Hahn and Carvalho}{Hahn and
  Carvalho}{2015}]{2015Hannp435448}
Hahn, P.~R. and C.~M. Carvalho (2015).
\newblock Decoupling shrinkage and selection in {B}ayesian linear models: a
  posterior summary perspective.
\newblock {\em Journal of the American Statistical Association\/}~{\em
  110\/}(509), 435--448.

\bibitem[\protect\citeauthoryear{Homser and Lemeshow}{Homser and
  Lemeshow}{1989}]{1989Homserp}
Homser, D.~W. and S.~Lemeshow (1989).
\newblock {\em Applied logistic regression}.
\newblock New York (NY): J Wiley and Sons.

\bibitem[\protect\citeauthoryear{Hurvich and Tsai}{Hurvich and
  Tsai}{1989}]{1989Hurvichp297307}
Hurvich, C.~M. and C.-L. Tsai (1989, June).
\newblock Regression and time series model selection in small samples.
\newblock {\em Biometrika\/}~{\em 76\/}(2), 297--307.

\bibitem[\protect\citeauthoryear{Kohavi}{Kohavi}{1995}]{1995Kohavip11371145}
Kohavi, R. (1995).
\newblock A study of cross-validation and bootstrap for accuracy estimation and
  model selection.
\newblock In {\em International Joint Conference on Artificial Intelligence},
  Volume~14, pp.\  1137--1145.

\bibitem[\protect\citeauthoryear{Kullback and Leibler}{Kullback and
  Leibler}{1951}]{1951Kullbackp7986}
Kullback, S. and R.~A. Leibler (1951).
\newblock On information and sufficiency.
\newblock {\em The Annals of Mathematical Statistics\/}~{\em 22\/}(1), 79--86.

\bibitem[\protect\citeauthoryear{Kyung, Gill, Ghosh, and Casella}{Kyung
  et~al.}{2010}]{2010Kyungp369412}
Kyung, M., J.~Gill, M.~Ghosh, and G.~Casella (2010).
\newblock Penalized regression, standard errors, and {B}ayesian {L}assos.
\newblock {\em Bayesian Analysis\/}~{\em 5\/}(2), 369--412.

\bibitem[\protect\citeauthoryear{Makalic and Schmidt}{Makalic and
  Schmidt}{2016a}]{2016EnespHDBRRWTBP}
Makalic, E. and D.~F. Schmidt (2016a).
\newblock High-dimensional {B}ayesian regularised regression with the bayesreg
  package.
\newblock {\em arXiv preprint arXiv:1611.06649\/}.

\bibitem[\protect\citeauthoryear{Makalic and Schmidt}{Makalic and
  Schmidt}{2016b}]{2016Enesp179182}
Makalic, E. and D.~F. Schmidt (2016b).
\newblock A simple sampler for the horseshoe estimator.
\newblock {\em IEEE Signal Processing Letters\/}~{\em 23\/}(1), 179--182.

\bibitem[\protect\citeauthoryear{McQuarrie and Tsai}{McQuarrie and
  Tsai}{1998}]{1998McQuarriep}
McQuarrie, A. D.~R. and C.-L. Tsai (1998).
\newblock {\em Regression and time series model selection}.
\newblock World Scientific.

\bibitem[\protect\citeauthoryear{Mitchell and Beauchamp}{Mitchell and
  Beauchamp}{1988}]{1988Mitchellp10231032}
Mitchell, T.~J. and J.~J. Beauchamp (1988).
\newblock Bayesian variable selection in linear regression.
\newblock {\em Journal of the American Statistical Association\/}~{\em
  83\/}(404), 1023--1032.

\bibitem[\protect\citeauthoryear{Park and Casella}{Park and
  Casella}{2008}]{2008Parkp681686}
Park, T. and G.~Casella (2008, June).
\newblock The {B}ayesian {L}asso.
\newblock {\em Journal of the American Statistical Association\/}~{\em
  103\/}(482), 681--686.

\bibitem[\protect\citeauthoryear{Pas, Kleijn, and Vaart}{Pas
  et~al.}{2014}]{2014vanderPasp25852618}
Pas, S. V.~D., B.~Kleijn, and A.~V.~D. Vaart (2014).
\newblock The horseshoe estimator: posterior concentration around nearly black
  vectors.
\newblock {\em Eectronic Journal of Statistics\/}~{\em 8\/}(2), 2585--2618.

\bibitem[\protect\citeauthoryear{Polson and Scott}{Polson and
  Scott}{2010}]{2010Polsonp501538}
Polson, N.~G. and J.~G. Scott (2010).
\newblock Shrink globally, act locally: sparse {B}ayesian regularization and
  prediction.
\newblock {\em Bayesian Statistics\/}~{\em 9}, 501--538.

\bibitem[\protect\citeauthoryear{Rissanen}{Rissanen}{1983}]{1983Rissanenp416431}
Rissanen, J. (1983).
\newblock A universal prior for integers and estimation by minimum description
  length.
\newblock {\em The Annals of Statistics\/}~{\em 11\/}(2), 416--431.

\bibitem[\protect\citeauthoryear{Schmidt and Makalic}{Schmidt and
  Makalic}{2009}]{2009Schmidtp312321}
Schmidt, D.~F. and E.~Makalic (2009).
\newblock {MML} invariant linear regression.
\newblock In {\em Australasian Joint Conference on Artificial Intelligence},
  pp.\  312--321.

\bibitem[\protect\citeauthoryear{Schwarz}{Schwarz}{1978}]{1978Schwarzp461464}
Schwarz, G. (1978).
\newblock Estimating the dimension of a model.
\newblock {\em The Annals of Statistics\/}~{\em 6\/}(2), 464--464.

\bibitem[\protect\citeauthoryear{Shao}{Shao}{1993}]{1993Shaop486}
Shao, J. (1993, 06).
\newblock Linear model selection by cross-validation.
\newblock {\em Journal of the American Statistical Association\/}~{\em
  88\/}(422), 486--494.

\bibitem[\protect\citeauthoryear{Stone}{Stone}{1974}]{1974Stonep111}
Stone, M. (1974).
\newblock Cross-validatory choice and assessment of statistical predictions.
\newblock {\em Journal of the Royal Statistical Society. Series B
  (Methodological)\/}~(2), 111--147.

\bibitem[\protect\citeauthoryear{Tibshirani}{Tibshirani}{1996}]{1996Tibshiranip267288}
Tibshirani, R. (1996).
\newblock Regression shrinkage and selection via the {L}asso.
\newblock {\em Journal of the Royal Statistical Soiety\/}~{\em 58\/}(1),
  267--288.

\bibitem[\protect\citeauthoryear{Wallace}{Wallace}{2005}]{2005WallaceSIIMML}
Wallace, C.~S. (2005).
\newblock Statistical and inductive inference by minimum message length.
\newblock {\em Springer Science \& Business Media\/}.

\bibitem[\protect\citeauthoryear{Xu, Schmidt, Makalic, Qian, and Hopper}{Xu
  et~al.}{2016}]{2016Xup229240}
Xu, Z., D.~F. Schmidt, E.~Makalic, G.~Qian, and J.~L. Hopper (2016).
\newblock Bayesian grouped horseshoe regression with application to additive
  models.
\newblock In {\em Australasian Joint Conference on Artificial Intelligence},
  pp.\  229--240. Springer.

\bibitem[\protect\citeauthoryear{Yuan and Lin}{Yuan and
  Lin}{2006}]{2006Yuanp4967}
Yuan, M. and Y.~Lin (2006).
\newblock Model selection and estimation in regression with grouped variables.
\newblock {\em Journal of the Royal Statistical Society: Series B (Statistical
  Methodology)\/}~{\em 68\/}(1), 49--67.

\bibitem[\protect\citeauthoryear{Zhang and Bondell}{Zhang and
  Bondell}{2017}]{2016Zhangp}
Zhang, Y. and H.~D. Bondell (2017).
\newblock Variable selection via penalized credible regions with
  {D}irichlet--{L}aplace global-local shrinkage priors.
\newblock {\em Bayesian Analysis\/}.

\bibitem[\protect\citeauthoryear{Zou}{Zou}{2006}]{2006Zoup14181429}
Zou, H. (2006).
\newblock The adaptive {L}asso and its oracle properties.
\newblock {\em Journal of the American Statistical Association\/}~{\em
  101\/}(476), 1418--1429.

\end{thebibliography}

\end{document}